\documentclass[aps,pre,nofootinbib]{revtex4}

\usepackage{amsmath,amsfonts,amssymb}
\usepackage{graphicx}

\def\s{\sigma} \def\us{\underline{\sigma}} \def\uT{\underline{T}}
\def\ut{\underline{\tau}} \def\E{\mathbb{E}} \def\X{{\cal X}}
\newcommand{\comment}[1]{}

\def\k{k}
\def\sat{\text{SAT}}
\def\unsat{\text{UNSAT}}
\def\xorsat{\text{XORSAT}}
\def\ksat{\k\text{-}\sat}
\def\kxorsat{\k\text{-}\xorsat}

\begin{document}

\title{
A review of the 
Statistical Mechanics approach to Random Optimization Problems
}
\author{Fabrizio Altarelli$^{\,1,2}$, R\'emi Monasson$^{\,2}$,
Guilhem Semerjian$^{\,2}$ and Francesco Zamponi$^{\,2}$}
\address{$^{1\,}$
Dipartimento di Fisica and CNR, Universit\`a di Roma La Sapienza,
P. A. Moro 2, 00185 Roma, Italy, \\
$^{2\,}$
LPTENS, Unit\'e Mixte de Recherche (UMR 8549) du CNRS et de l'ENS,
associ\'ee \`a l'UPMC Univ Paris 06, 24 Rue Lhomond, 75231 
Paris Cedex 05, France.
}

\begin{abstract}
We review the connection between statistical mechanics and the analysis
of random optimization problems, with particular emphasis on the random
$k$-SAT problem. 
We discuss and characterize the different phase transitions that are
met in these problems, starting from basic concepts.
We also discuss how statistical mechanics methods can be used to investigate
the behavior of local search and decimation based algorithms. \\
{\it This paper has been written as a contribution to the ``Handbook of
Satisfiability'' to be published in 2008 by IOS press.}
\end{abstract}

\maketitle

\section{Introduction}

The connection between the statistical physics of disordered systems
and optimization problems in computer science dates back from twenty
years at least~\cite{Beyond}. In combinatorial optimization one 
is given a cost function (the length of a tour in the traveling salesman
problem (TSP), the number of violated constraints in constraint 
satisfaction problems,~\dots)
over a set of variables and looks for the minimal cost over
an allowed range for those variables. Finding the true minimum
may be complicated, and requires bigger and bigger computational
efforts as the number of variables to be minimized over
increases~\cite{Pa83}. 
Statistical physics is at first sight very
different. The scope is to deduce the macroscopic, that is, global
properties of a physical system, for instance a gas, a liquid or a solid,  
from the knowledge of the energetic interactions of its elementary
components (molecules, atoms or ions). However, at very
low temperature, these elementary components are essentially forced to
occupy the spatial conformation minimizing the global energy
of the system. Hence low temperature statistical physics can
be seen as the search for minimizing a cost function whose expression
reflects the laws of Nature or, more humbly, the degree of accuracy retained
in its description. This problem is generally not difficult to solve
for non disordered systems where 
the lowest energy conformation are crystals in which components are
regularly spaced from each other. Yet the presence of disorder,
e.g. impurities, makes the problem very difficult and finding the
conformation with minimal energy is a true optimization problem. 

At the beginning of the eighties, 
following the works of G. Parisi and others on systems called spin
glasses~\cite{Beyond}, 
important progresses were made in the statistical physics of 
disordered systems. Those progresses made possible the quantitative study of 
the properties of systems given some distribution of the disorder 
(for instance the location of impurities) such as the 
average minimal energy and its fluctuations.
The application to optimization problems was natural and led to
beautiful studies on (among others) the average
properties of the minimal tour length in the TSP, 
the minimal cost in Bipartite Matching, for some specific
instance distributions~\cite{Beyond}. Unfortunately 
statistical physicists and computer scientists did not establish 
close ties on a large scale at that time. The reason could 
have been of methodological nature~\cite{Fu85}. 
While physicists were making statistical
statements, true for a given distribution of inputs, computer scientists were
rather interested in solving one (or several) particular instances of a
problem. The focus was thus on efficient ways to do so, that is, requiring a
computational effort growing not too quickly with the number of data
defining the instance.  Knowing precisely the typical properties for
a given, academic distribution of instances did not help much to
solve practical cases.

At the beginning of the nineties practitionners in artificial
intelligence realized that classes of random constraint satisfaction
problems used as artificial benchmarks for search algorithms exhibited
abrupt changes of behaviour when some control parameter were finely
tuned~\cite{transition_exp}. The most celebrated example was random
$\k$-Satisfiability, where one looks for a solution to a set of random
logical constraints over a set of Boolean variables. It appeared that,
for large sets of variables, there was a critical value of the number
of constraints per variable below which there almost surely existed
solutions, and above which solutions were absent. An important feature
was that the performances of known search algorithms drastically worsened 
in the vicinity of this critical ratio. In addition to its intrinsic
mathematical interest the random $\k$-SAT problem was therefore worth to
be studied for `practical' reasons. 

This critical phenomenon, strongly reminiscent of phase transitions in
condensed matter physics, led to a revival of the research at the 
interface between statistical physics and computer science, which is still
very active. The purpose of the present review is to introduce the non
physicist reader to some concepts required to understand the
literature in the field and to present some major results.
We shall in
particular discuss the refined picture of the satisfiable phase put
forward in statistical mechanics studies and the algorithmic approach
(Survey Propagation, an extension of Belief Propagation used in
communication theory and statistical inference) this picture suggested. 

While the presentation will mostly focus on the $\k$-Satisfiability
problem (with random constraints) we will occasionally discuss
another computational problem, namely, linear systems of Boolean
equations.  A good reason to do so is that this problem exhibits some
essential features encountered in random $\k$-Satisfiability, while
being technically simpler to study. In addition it is closely related
to error-correcting codes in communication theory.

The chapter is divided into four main parts. In Section~\ref{sec_basic} 
we present the
basic statistical physics concepts necessary to understand the onset
of phase transitions, and to characterize the nature of the
phases. Those are illustrated on a simple example of decision problem,
the so-called perceptron problem. In Section~\ref{sec:phase_transitions} 
we review the scenario of the various phase transitions taking place in
random $\k$-SAT.
Section~\ref{sec_localsearch} and~\ref{sec_decimation}
present the techniques used to study various type of algorithms in optimization
(local search, backtracking procedures, message
passing algorithms). We end up with some conclusive remarks in 
Sec.~\ref{sec_conclu}.

\section{Phase Transitions: Basic Concepts and Illustration}

\label{sec_basic}
\index{phase transition}

\subsection{A simple decision problem with a phase transition: the
  continuous perceptron}

For pedagogical reasons we first discuss a simple example exhibiting
several important features we shall define more formally in the next
subsection.  Consider $M$ points $\uT^1,\dots,\uT^M$ of the
$N$-dimensional space $\mathbb{R}^N$, their coordinates being denoted
$\uT^a=(T_1^a,\dots,T_N^a)$. The continuous perceptron problem
consists in deciding the existence of a vector $\us \in \mathbb{R}^N$
which has a positive scalar product with all vectors linking the
origin of $\mathbb{R}^N$ to the $\uT$'s,
\begin{equation}\label{question}
\us \cdot \uT^a \equiv \sum _{i=1} ^N \sigma_i \; T_i^a > 0 \ , \qquad
\forall\, a=1,\ldots,M \quad ,
\end{equation}
or in other words determining whether the $M$ points belong to the
same half-space. The term continuous in the name of the problem
emphasizes the domain $\mathbb{R}^N$ of the variable $\us$. This makes
the problem polynomial from worst-case complexity point of view 
\cite{revue-perceptron}.

\begin{figure}
\includegraphics[width=7cm]{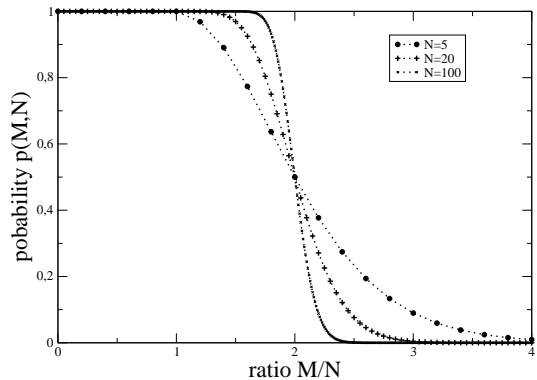}
\caption{Probability $P(N,M)$ that $M$ random points on the
$N$-dimensional unit hypersphere are located in the same half-space.
Symbols correspond to Cover's exact result~\cite{Cover}, see
Eq.~(\ref{percep1}), lines serve as guides to the eye.}
\label{fig-proba}
\end{figure}

Suppose now that the points are chosen independently, identically,
uniformly on the unit hypersphere, and call
\begin{eqnarray}
P (N,M) &=& \mbox{Probability that a set of $M$ randomly chosen points} 
\nonumber \\&&\mbox{belong to the same half-space.}
\nonumber
\end{eqnarray}
This quantity can be computed exactly~\cite{Cover} (see also Chapter
5.7 of~\cite{revue-perceptron}) and is plotted in Fig.~\ref{fig-proba}
as a function of the ratio $\alpha = M/N$ for increasing sizes
$N=5,20,100$.  Obviously $P$ is a decreasing function of the number
$M$ of points for a given size $N$: increasing the number of
constraints can only make more difficult the simultaneous satisfaction
of all of them. More surprisingly, the figure suggests that, in the
large size limit $N\to\infty$, the probability $P$ reaches a limiting
value 0 or 1 depending on whether the ratio $\alpha$ lies,
respectively, above or below some `critical' value $\alpha_{\rm
  s}=2$. This is confirmed by the analytical expression of $P$
obtained in~\cite{Cover},
\begin{equation}\label{percep1}
P(N,M)=\frac{1}{2^{M-1}} \sum _{i=0}^{\min(N-1,M-1)} \binom{M-1}{i} \
,
\end{equation}
from which one can easily show that, indeed,
\begin{equation}\label{percep_trans}
\lim _{N\to\infty} P(N,M= N \alpha) =
\begin{cases}
1 & \mbox{if} \ \alpha < \alpha_{\rm s} \\ 0 & \mbox{if} \ \alpha >
\alpha_{\rm s}
\end{cases}
\ , \qquad \mbox{with} \ \alpha_{\rm s}=2 \ .
\end{equation}

Actually the analytical expression of $P$ allows to describe more
accurately the drop in the probability as $\alpha$ increases. To this
aim we make a zoom on the transition region $M \approx N \alpha_{\rm
s} $ and find from (\ref{percep1}) that
\begin{equation}\label{percep_fss}
\underset{N \to \infty}{\lim} P(N, M = N \alpha_{\rm s} (1+ \lambda \,
N^{-1/2} )\,)= \int_{\lambda \sqrt{2}}^{\infty} \frac{dx}
{\sqrt{2\pi}}\, e^{-x^2/2} \ .
\end{equation}
As it should the limits $\lambda \to \pm \infty$ gives back the coarse
description of Eq.~(\ref{percep_trans})

\subsection{Generic definitions}
\label{sec_gendef}
We now put this simple example in a broader perspective and introduce
some generic concepts that it illustrates, along with the definitions
of the problems studied in the following.

\begin{itemize}

\item[$\bullet$] Constraint Satisfaction Problem (CSP)

A CSP is a decision problem where an assignment (or configuration) of
$N$ variables $\us=(\s_1,\dots,\s_N) \in \X^N$ is required to
simultaneously satisfy $M$ constraints. In the continuous perceptron
the domain of $\us$ is $\mathbb{R}^N$ and the constraints impose the
positivity of the scalar products (\ref{question}). The instance of
the CSP, also called formula in the following, is said satisfiable if
there exists a solution (an assignment of $\us$ fulfilling all the
constraints). The $\ksat$ problem is a boolean CSP ($\X = \{ {\rm
True}, {\rm False} \}$) where each constraint (clause) is the
disjunction (logical OR) of $\k$ literals (a variable or its negation).
Similarly in $\kxorsat$ the literals are combined by an eXclusive OR
operation, or equivalently an addition modulo 2 of $0/1$ boolean
variables is required to take a given value.  The worst-case
complexities of these two problems are very different ($\kxorsat$ is
in the P complexity class for any $\k$ while $\ksat$ is NP-complete
for any $\k \ge 3$), yet for the issues of this review we shall see
that they present a lot of similarities.  In the following we use the
statistical mechanics convention and represent boolean variables by
Ising spins, $\X=\{-1,+1\}$. A $\ksat$ clause will be defined by $\k$
indices $i_1,\dots,i_\k \in [1,N]$ and $\k$ values
$J_{i_1},\dots,J_{i_\k} = \pm 1$, such that the clause is unsatisfied
by the assignment $\us$ if and only if $\s_{i_j} = J_{i_j} \ \ \forall
j \in [1,\k]$.  A $\kxorsat$ clause is satisfied if the product of the
spins is equal to a fixed value, $\s_{i_1} \dots \s_{i_\k} = J$.

\item[$\bullet$] random Constraint Satisfaction Problem (rCSP)

The set of instances of most CSP can be turned in a probabilistic
space by defining a distribution over its constraints, as was done in
the perceptron case by drawing the vertices $\uT^a$ uniformly on the
hypersphere.  The random $\ksat$ formulas considered in the following
are obtained by choosing for each clause $a$ independently a $\k$-uplet
of distinct indices $i_1^a,\dots,i_\k^a$ uniformly over the
$\binom{N}{\k}$ possible ones, and negating or not the corresponding
literals ($J_i^a = \pm 1$) with equal probability one-half. The
indices of random $\xorsat$ formulas are chosen similarly, with the
constant $J^a = \pm 1$ uniformly.

\item[$\bullet$] thermodynamic limit and phase transitions

These two terms are the physics jargon for, respectively, the large
size limit ($N \to \infty$) and for threshold phenomena as stated for
instance in (\ref{percep_trans}). In the thermodynamic limit the
typical behavior of physical systems is controlled by a small number
of parameters, for instance the temperature and pressure of a gas. At
a phase transition these systems are drastically altered by a tiny
change of a control parameter, think for instance at what happens to
water when its temperature crosses $100 \, ^\mathrm{o}$C . 
This critical value of
the temperature separates two qualitatively distinct phases, liquid
and gaseous. For random CSPs the role of control parameter is usually
played by the ratio of constraints per variable, $\alpha = M /N$, kept
constant in the thermodynamic limit. Eq.~(\ref{percep_trans})
describes a satisfiability transition for the continuous perceptron,
the critical value $\alpha_{\rm s}=2$ separating a satisfiable phase
at low $\alpha$ where instances typically have solutions to a phase
where they typically do not. Typically is used here as a synonym for
with high probability, i.e. with a probability which goes to one in
the thermodynamic limit.

\item[$\bullet$] Finite Size Scaling (FSS) \index{finite size scaling}

The refined description of the neighborhood of the critical value of
$\alpha$ provided by (\ref{percep_fss}) is known as a finite size
scaling relation.  More generally the finite size scaling hypothesis
for a threshold phenomenon takes the form
\begin{equation} \label{hyposca}
\underset{N \to \infty}{\lim} P(N, M = N \alpha_{\rm s} (1+ \lambda \,
N^{-1/\nu} )\,)= {\cal F}(\lambda) \ ,
\end{equation}
where $\nu$ is called the FSS exponent ($2$ for the continuous
perceptron) and the scaling function ${\cal F}(\lambda)$ has limits
$1$ and $0$ at respectively $-\infty$ and $+\infty$. This means that,
for a large but finite size $N$, the transition window for the values
of $M/N$ where the probability drops from $1 - \epsilon$ down to
$\epsilon$ is, for arbitrary small $\epsilon$, of width
$N^{-1/\nu}$. Results of this flavour are familiar in the study of random 
graphs~\cite{random_graphs}; for instance the appearance of a giant 
component containing a finite fraction of the vertices of an Erd\"os-R\'enyi
random graph happens on a window of width $N^{-1/3}$ on the average 
connectivity.
FSS relations are important, not only from the
theoretical point of view, but also for practical applications. Indeed
numerical experiments are always performed on finite-size instances
while theoretical predictions on phase transitions are usually true in
the $N\to\infty$ limit. Finite-size scaling relations help to bridge
the gap between the two.  We shall review some FSS results in
Sec.~\ref{sec_review_FSS}.
\end{itemize}

Let us emphasize that random $\ksat$, and other random CSP, are
expected to share some features of the continuous perceptron model,
for instance the existence of a satisfiability threshold, but of course
not its extreme analytical simplicity. In fact, despite an intensive
research activity, the mere existence of a satisfiability threshold
for random $\sat$ formulas remains a (widely accepted) conjecture. A
significant achievement towards the resolution of the conjecture was
the proof by Friedgut of the existence of a non-uniform sharp
threshold~\cite{Friedgut}. There exists also
upper~\cite{transition_ub} and lower~\cite{transition_lb} bounds on
the possible location of this putative threshold, which become almost
tight for large values of $\k$~\cite{transition_largek}. We refer the
reader to the chapter~\cite{chapter_randomsat} of this volume for more
details on these issues. This difficulty to obtain tight results with
the currently available rigorous techniques is a motivation for the use
of heuristic statistical mechanics methods, that provide intuitions on
why the standard mathematical ones run into trouble and how to amend
them. In the recent years important results first conjectured by
physicists were indeed rigorously proven.  Before describing in some
generality the statistical mechanics approach, it is instructive to
study a simple variation of the perceptron model for which the basic
probabilistic techniques become inefficient.

\subsection{The perceptron problem continued: binary variables}
\label{bper}

The binary perceptron problem consists in looking for solutions of
(\ref{question}) on the hypercube i.e. the domain of the variable
$\us$ is $\X^N = \{ -1 , +1 \}^N$ instead of $\mathbb{R}^N$. This
decision problem is NP-complete. 
Unfortunately Cover's calculation~\cite{Cover} cannot be extended to
this case, though it is natural to expect a similar satisfiability
threshold phenomenon at an a priori distinct value $\alpha_{\rm
s}$. Let us first try to study this point with basic probabilistic
tools, namely the first and second moment method~\cite{2nd_moment}.
The former is an application of the Markov inequality,
\begin{equation}
{\rm Prob}[Z > 0] \le \E[Z] \ ,
\label{eq_Markov_inequality}
\end{equation}
valid for positive integer valued random variables $Z$. We shall use
it taking for $Z$ the number of solutions of (\ref{question}),
\begin{equation}
Z = \sum_{\us \in \X^N} \prod_{a=1} ^M \theta (\us \cdot \uT^a) \ ,
\end{equation}
where $\theta(x)=1$ if $x>0$, $0$ if $x\le 0$. The expectation value
of the number of solutions is easily computed,
\begin{equation}
\E [ Z ] = 2^N \times 2^{-M} = e^{N\, G_1} \quad \mbox{with} \quad G_1
= (1-\alpha) \ln 2 \ ,
\end{equation}
and vanishes when $N\to\infty$ if $\alpha >1$. Hence, from Markov's
inequality (\ref{eq_Markov_inequality}), with high probability
constraints (\ref{question}) have no solution on the hypercube when
the ratio $\alpha$ exceeds unity: if the threshold $\alpha_{\rm s}$
exists, it must satisfy the bound $\alpha_{\rm s} \le 1$. One can look
for a lower bound to $\alpha_{\rm s}$ using the second moment method,
relying on the inequality~\cite{2nd_moment}
\begin{equation}
\frac{\E[Z]^2}{\E[Z^2]} \le {\rm Prob}[Z > 0] \ .
\label{eq_inequality_2}
\end{equation}
The expectation value of the squared number of solutions reads
\begin{equation}\label{mom2}
\E [Z^2] = \sum _{\us,\us'} \left( \E [\theta (\us \cdot \uT)
\;\theta (\us' \cdot \uT)] \right)^M
\end{equation}
since the vertices $\uT^a$ are chosen independently of each other.
The expectation value on the right hand side of the above expression
is simply the probability that the vector pointing to a randomly
chosen vertex, $\uT$, has positive scalar product with both vectors
$\us,\us'$. Elementary geometrical considerations reveal that
\begin{equation}
\E [\theta (\us \cdot \uT) \;\theta (\us' \cdot \uT)] = \frac
1{2\pi} \left( \pi - \varphi( \us,\us') \right)
\end{equation}
where $\varphi$ is the relative angle between the two vectors. This angle
can be alternatively parametrized by the overlap between $\us$ and $\us'$,
i.e. the normalized scalar product,
\begin{equation}
q=\frac 1N \sum _{i=1}^N \sigma_i \, \sigma'_i  = 1 -2 \frac{1}{N}
\sum_{i=1}^N \mathbb{I}(\s_i \neq \s'_i ) \ .
\label{eq_traduc_Hamm}
\end{equation}
The last expression, in which $\mathbb{I}(E)$ denotes the indicator function
of the event $E$, reveals the traduction between the concept of overlap
and the more traditional Hamming distance.
The sum over vectors in (\ref{mom2}) can then be replaced by a sum
over overlap values with appropriate combinatorial coefficients
counting the number of pairs of vectors at a given overlap. The
outcome is
\begin{equation}
\E [Z^2] = 2^N\sum _{q=-1,-1+\frac 2N, -1+\frac 4N , \ldots ,1}
\binom{N}{N\left(\frac{1+q}2\right)} \ \left(\frac 12 - \frac 1 {2\pi}
\;\mbox{Arcos} \; q \right)^M \ .
\end{equation}
In the large $N$ limit we can estimate this sum with the Laplace
method,
\begin{equation}
\lim _{N\to\infty} \frac 1N \ln \E [Z^2] = \max _{-1< q < 1} G_2(q) \
,
\end{equation}
where
\begin{eqnarray}
G_2(q) = \ln 2 &-&\left(\frac{1+q}2\right) \ln
\left(\frac{1+q}2\right) - \left(\frac{1-q}2\right) \ln
\left(\frac{1-q}2\right) \nonumber \\
&+& \alpha\; \ln \left(\frac 12 - \frac 1
     {2\pi} \; \mbox{Arcos} q \right)\ .\label{defg}
\end{eqnarray}
Two conclusions can be drawn from the above calculation:
\begin{itemize}
\item no useful lower bound to $\alpha_{\rm s}$ can be obtained from
such a direct application of the second moment method.  Indeed,
maximization of $G_2$ (\ref{defg}) over $q$ shows that $\E [Z^2] \gg
(\E[Z])^2$ when $N$ diverges, whenever $\alpha >0$, and in consequence
the left hand side of (\ref{eq_inequality_2}) vanishes.  A possible
scenario which explains this absence of concentration of the number of
solutions is the following. As shown by the moment calculation the
natural scaling of $Z$ is exponentially large in $N$ (as is the total
configuration space $\X^N$). We shall thus denote $s = (\ln Z)/N$ the
random variable of order one counting the log degeneracy of the
solutions. Suppose $s$ follows a large deviation principle~\cite{ldp}
that we state in a very rough way as ${\rm Prob}[s] \approx \exp[N
L(s)]$, with $L(s)$ a negative rate function, assumed for simplicity
to be concave. Then the moments of $Z$ are given, at the leading
exponential order, by
\begin{equation}
\lim_{N \to \infty} \frac{1}{N} \ln \E[Z^n] = \max_s [ L(s) + n s] \
,
\end{equation}
and are controlled by the values of $s$ such that $L'(s)=-n$. The
moments of larger and larger order $n$ are thus dominated by the
contribution of rarer and rarer instances with larger and larger
numbers of solutions. On the contrary the typical value of the number
of solutions is given by the maximum of $L$, reached in a value we
denote $s_{\rm g}(\alpha)$: with high probability when $N\to\infty$,
$Z$ is comprised between $e^{N(s_{\rm g}(\alpha)-\epsilon)}$ and
$e^{N(s_{\rm g}(\alpha)+\epsilon)}$, for any $\epsilon >0$. From this
reasoning it appears that the relevant quantity to be computed is
\begin{equation} \label{defsg0}
s_{\rm g}(\alpha) = \lim_{N\to\infty} \frac{1}{N} \E [ \ln Z ] =
\lim _{N\to\infty} \lim_{n\to 0} \frac 1{n\,N} \ln \E [Z^n] \ .
\end{equation}
This idea of computing moments of vanishing order is known in
statistical mechanics as the replica\footnote{The vocable replicas
comes from the presence of $n$ copies of the vector $\us$ in the
calculation of $Z^n$ (see the $n=2$ case in formula (\ref{mom2})).}
method~\cite{Beyond}. \index{replica method}
Its non-rigorous implementation consists in
determining the moments of integer order $n$, which are then 
continued towards $n=0$. The outcome of such a computation for the
binary perceptron problem reads~\cite{kra89}
\begin{eqnarray}\label{solution_sg}
s_{\rm g}(\alpha) = \max _{q,\hat q} \bigg\{ &-&\frac 12 q (1-\hat q) +
\int _{-\infty}^\infty Dz \ln (2 \cosh(z \sqrt{\hat q})) \\
&+& \alpha
\int _{-\infty}^\infty Dz \ln \left[\int _{z \sqrt{q/(1-q)}} ^\infty
Dy \right] \bigg\} \ ,\nonumber
\end{eqnarray}
where $Dz\equiv dz\; e^{-z^2/2}/\sqrt{2\pi}$. The entropy $s_{\rm
g}(\alpha)$ is a decreasing function of $\alpha$, which vanishes in
$\alpha_{\rm s} \simeq 0.833$. Numerical experiments support this
value for the critical ratio of the satisfiable/unsatisfiable phase
transition.

\item the calculation of the second moment is naturally related to the
determination of the value of the overlap $q$ between pairs of
solutions (or equivalently their Hamming distance, recall 
Eq.~(\ref{eq_traduc_Hamm})). This conclusion extends to the calculation of 
the $n^{th}$
moment for any integer $n$, and to the $n\to 0$ limit. The value of
$q$ maximizing the r.h.s. of (\ref{solution_sg}), $q^*(\alpha)$,
represents the average overlap between two solutions of the same set
of constraints (\ref{question}). Actually the distribution of overlaps
is highly concentrated in the large $N$ limit around
$q^*(\alpha)$, in other words the (reduced) Hamming distance between
two solutions is, with high probability, equal to
$d^* (\alpha) = (1-q^*(\alpha))/2$. This distance $d^*(\alpha)$
ranges from $\frac 12$ for $\alpha=0$ to $\simeq \frac 14$ at
$\alpha=\alpha_{\rm s}$. Slightly below the critical ratio solutions
are still far away from each other on the hypercube\footnote{This
situation is very different from the continuous perceptron case, where
the typical overlap $q^*(\alpha)$ reaches one when $\alpha$ tends to
2: a single solution is left right at the critical ratio.}.
\end{itemize}

Note that the perceptron problem is not as far as it could seem from
the main subject of this review.  There exists indeed a natural
mapping between the binary perceptron problem and $\ksat$. Assume the
vertices $\uT$ of the perceptron problem, instead of being drawn on
the hypersphere, have coordinates that can take three values:
$T_i=-1,0,1$. Consider now a $\ksat$ formula $F$. To each clause $a$
of $F$ we associate the vertex $\uT^a$ with coordinates $T_i^a=-J_i^a$
if variable $i$ appears in clause $a$, $0$ otherwise.  Of course
$\sum_i |T_i^a|=\k$: exactly $\k$ coordinates have non zero values for
each vertex. Then replace condition (\ref{question}) with
\begin{equation}\label{question2}
\sum _{i=1} ^N \sigma_i \; T_i^a > -(\k-1) \ , \qquad \forall\,
a=1,\ldots,M \ .
\end{equation}
The scalar product is not required to be positive any longer, but to
be larger than $-(\k-1)$. It is an easy check that the perceptron
problem admits a solution on the hypercube ($\sigma _i=\pm 1$) if and
only if $F$ is satisfiable. While in the binary perceptron model all
coordinates are non-vanishing, only a finite number of them take non
zero values in $\ksat$. For this reason $\ksat$ is called a diluted
model in statistical physics.

Also the direct application of the second moment method fails for the
random $\ksat$ problem; yet a refined version of it was used
in~\cite{transition_largek}, which leads to asymptotically (at large
$\k$) tight bounds on the location of the satisfiability threshold.

\subsection{From random CSP to statistical mechanics of disordered systems}

The binary perceptron example taught us that the number of solutions
$Z$ of a satisfiable random CSP usually scales exponentially with the
size of the problem, with large fluctuations that prevent the direct
use of standard moment methods. This led us to the
introduction of the quenched entropy, as defined in
(\ref{defsg0}). The computation techniques used to obtain
(\ref{solution_sg}) were in fact developed in an apparently different
field, the statistical mechanics of disordered systems~\cite{Beyond}.

Let us review some basic concepts of statistical
mechanics (for introductory books see for example~\cite{statmech1,statmech2}). 
A physical system can be modeled by a
space of configuration $\us \in \X^N$, on which is defined an energy
function $E(\us)$. For instance usual magnets are described by Ising
spins $\s_i=\pm 1$, the energy being minimized when
adjacent spins take the same value. The equilibrium properties of a
physical system at temperature $T$ are given by the Gibbs-Boltzmann
probability measure on $\X^N$,
\begin{equation}
\mu(\us) = \frac{1}{Z} \exp[-\beta E(\us) ] \ ,
\label{Gibbs}
\end{equation}
where the inverse temperature $\beta$ equals $1/T$ and $Z$ is a
normalization called partition function. The energy function $E$ has a
natural scaling, linear in the number $N$ of variables (such a
quantity is said to be extensive). In consequence in the thermodynamic
limit the Gibbs-Boltzmann measure concentrates on configurations with
a given energy density ($e=E/N$), which depends on the conjugated
parameter $\beta$. The number of such configurations is usually
exponentially large, $\approx \exp[N s]$, with $s$ called the entropy
density. The partition function is thus dominated by the contribution
of these configurations, hence $\lim (\ln Z /N) = s - \beta e$.

In the above presentation we supposed the energy to be a simple, known
function of the configurations. In fact some magnetic compounds,
called spin-glasses, are intrinsically disordered on a microscopic
scale. This means that there is no hope in describing exactly their
microscopic details, but that one should rather assume their energy to
be itself a random function with a known distribution. Hopefully in
the thermodynamic limit the fluctuations of the thermodynamic
observables as the energy and entropy density vanish, hence the
properties of a typical sample will be closely described by the
average (over the distribution of the energy function) of the entropy
and energy density.

The random CSPs fit naturally in this line of research. The energy
function $E(\us)$ of a CSP is defined as the number of constraints
violated by the assignment $\us$, in other words this is the cost
function to be minimized in the associated optimization problem
(MAX$\sat$ for instance).  Moreover the distribution of random instances
of CSP is the counterpart of the distribution over the microscopic
description of a disordered solid.  The study of the optimal
configurations of a CSP, and in particular the characterization of a
satisfiability phase transition, is achieved by taking the $\beta \to
\infty$ limit. Indeed, when this parameter increases (or equivalently
the temperature goes to 0), the law (\ref{Gibbs}) favors the lowest
energy configurations. In particular if the formula is satisfiable
$\mu$ becomes the uniform measures over the solutions. Two important
features of the formula can be deduced from the behavior of $Z$ at
large $\beta$: the ground-state energy $E_{\rm g} = \min_{\us}
E(\us)$, which indicates how good are the optimal configurations, and
the ground state entropy $S_{\rm g} = \ln (|\{ \us \ : \ E(\us) =
E_{\rm g} \} | )$, which counts the degeneracy of these optimal
configurations. The satisfiability of a formula is equivalent to its
ground-state energy being equal to 0. In the large $N$ limit these two
thermodynamic quantities are supposed to concentrate around their mean values 
(this is proven for $E$ in~\cite{purelit}), we thus introduce the associated 
typical densities,
\begin{equation} \label{defsg2}
e_{\rm g}(\alpha) = \lim_{N \to \infty } \frac{1}{N} \E [E_{\rm g}] \ , \qquad 
s_{\rm g}(\alpha) = \lim_{N \to \infty } \frac{1}{N} \E [S_{\rm g}] \ .
\end{equation}
Notice that formula (\ref{defsg2}) coincides with (\ref{defsg0}) in
the satisfiable phase (where the ground state energy vanishes).

Some criteria are needed to relate these thermodynamic quantities to
the (presumed to exist) satisfiability threshold $\alpha_{\rm s}$. A
first approach, used for instance in~\cite{MoZe}, consists in locating
it as the point where the ground-state energy density $e_{\rm g}$
becomes positive. The assumption underlying this reasoning is the
absence of an intermediate, typically $\unsat$ regime, with a
sub-extensive positive $E_{\rm g}$.  In the discussion of the binary
perceptron we used another criterion, namely we recognized
$\alpha_{\rm s}$ by the cancellation of the ground-state entropy
density. This argument will be true if the typical number of solutions
vanishes continuously at $\alpha_{\rm s}$. It is easy to realize that
this is not the case for random $\ksat$: at any finite value of
$\alpha$ a finite fraction $\exp[-\alpha \k]$ of the variables do not
appear in any clause, which leads to a trivial lower bound $(\ln 2)
\exp[-\alpha \k]$ on $s_{\rm g}$. This quantity is thus finite at the
transition, a large number of solutions disappear suddenly at
$\alpha_{\rm s}$. Even if it is wrong, the criterion $s_g(\alpha) = 0$ 
for the determination of the satisfiability transition is instructive for two
reasons. First, it becomes asymptotically correct at large $\k$ (free
variables are very rare in this limit), this is why it works for the
binary perceptron of Section \ref{bper} (which is, as we have seen,
close to $\ksat$ with $\k$ of order $N$). Second, it will reappear below
in a refined version: we shall indeed decompose the entropy in two
qualitatively distinct contributions, one of the two being indeed
vanishing at the satisfiability transition.

\section{Phase transitions in random CSPs}
\label{sec:phase_transitions}

\subsection{The clustering phenomenon}
\label{sec_clustering}
\index{phase transition!clustering transition}

We have seen that the statistical physics approach to the perceptron
problem naturally provided us with information about the geometry of
the space of its solutions. Maybe one of the most important
contribution of physicists to the field of random CSP was to suggest
the presence of further phase transitions in the satisfiable regime
$\alpha < \alpha_{\rm s}$, affecting qualitatively the geometry
(structure) of the set of solutions~\cite{BiMoWe,MeZe,KrMoRiSeZd}.

This subset of the configuration space is indeed thought to break down
into ``clusters'' in a part of the satisfiable phase, $\alpha \in
[\alpha_{\rm d}, \alpha_{\rm s}]$, $\alpha_{\rm d}$ being the
threshold value for the clustering transition. Clusters are meant as a
partition of the set of solutions having certain properties listed
below.  Each cluster contains an exponential number of solutions,
$\exp[N s_{\rm int}]$, and the clusters are themselves exponentially
numerous, $\exp [N \Sigma]$. The total entropy density thus decomposes
into the sum of $s_{\rm int}$, the internal entropy of the clusters
and $\Sigma$, encoding the degeneracy of these clusters, usually
termed complexity in this context.  Furthermore, solutions inside a
given cluster should be well-connected, while two solutions of
distinct clusters are well-separated. A possible definition for these
notions is the following. Suppose $\us$ and $\ut$ are two solutions of
a given cluster. Then one can construct a path
$(\us=\us_0,\us_1,\dots,\us_{n-1},\us_n=\ut)$ where
any two
successive $\us_i$ are separated by a sub-extensive Hamming
distance. On the contrary such a path does not exist if $\us$ and
$\ut$ belong to two distinct clusters.  Clustered configuration spaces
as described above have been often encountered in various contexts,
e.g. neural networks~\cite{neural_networks} and mean-field spin
glasses~\cite{pspin}. A vast body of involved, yet non-rigorous,
analytical techniques~\cite{Beyond} have been developed in the field
of statistical mechanics of disordered systems to tackle such
situations, some of them having been justified
rigorously~\cite{Talagrand_book,PaTa,FrLe}.
In this
literature clusters appear under the name of ``pure states'', or
``lumps'' (see for instance the chapter 6 of~\cite{Talagrand_book} for
a rigorous definition and proof of existence in a related model).  As
we shall explain in a few lines, this clustering phenomenon has been
demonstrated rigorously in the case of random $\xorsat$
instances~\cite{xor_1,xor_2}. For random $\sat$ instances, where in fact
the detailed picture of the satisfiable phase is thought to be 
richer~\cite{KrMoRiSeZd},
there are some rigorous 
results~\cite{clus_rig_xsat1,clus_rig_xsat2,clus_rig_Fede} on
the existence of clusters for large enough $\k$.

\subsection{Phase transitions in random $\xorsat$}

Consider an instance $F$ of the $\xorsat$ problem~\cite{xor_replica},
i.e. a list of $M$ linear equations each involving $\k$ out of $N$ boolean
variables, where the additions are computed modulo 2. The study
performed in~\cite{xor_1,xor_2} provides a detailed picture of the
clustering and satisfiability transition sketched above. A crucial point
is the construction of a core subformula according to the following
algorithm. Let us denote $F_0 = F$ the initial set of equations, and
$V_0$ the set of variables which appear in at least one equation of
$F_0$.  A sequence $F_T,V_T$ is constructed recursively: if there are
no variables in $V_T$ which appear in exactly one equation of $F_T$
the algorithm stops.  Otherwise one of these ``leaf variables''
$\sigma_i$ is chosen arbitrarily, $F_{T+1}$ is constructed from $F_T$
by removing the unique equation in which $\sigma_i$ appeared, and
$V_{T+1}$ is defined as the set of variables which appear at least
once in $F_{T+1}$. Let us call $T_*$ the number of steps performed
before the algorithm stops, and $F'=F_{T_*}$, $V'=V_{T_*}$ the
remaining clauses and variables. Note first that despite the
arbitrariness in the choice of the removed leaves, the output
subformula $F'$ is unambiguously determined by $F$. Indeed, $F'$ can be
defined as the maximal (in the inclusion sense) subformula in which
all present variables have a minimal occurrence number of 2, and is
thus unique. In graph theoretic terminology $F'$ is the 2-core of $F$,
the $q$-core of hypergraphs being a generalization of the more
familiar notion on graphs, thoroughly studied in random graph
ensembles in~\cite{q_core_graphs}. Extending this study, relying on
the approximability of this leaf removal process by differential
equations~\cite{diff_eq}, it was shown in~\cite{xor_1,xor_2} that
there is a threshold phenomenon at $\alpha_{\rm d}(\k)$. For $\alpha <
\alpha_{\rm d}$ the 2-core $F'$ is, with high probability, empty, whereas
it contains a finite fraction of the variables and equations for
$\alpha > \alpha_{\rm d}$. $\alpha_{\rm d}$ is easily determined
numerically: it is the smallest value of $\alpha$ such that the
equation $x=1-\exp[-\alpha \k x^{\k-1}]$ has a non-trivial solution in
$(0,1]$. 

\begin{figure}
\includegraphics[width=9cm]{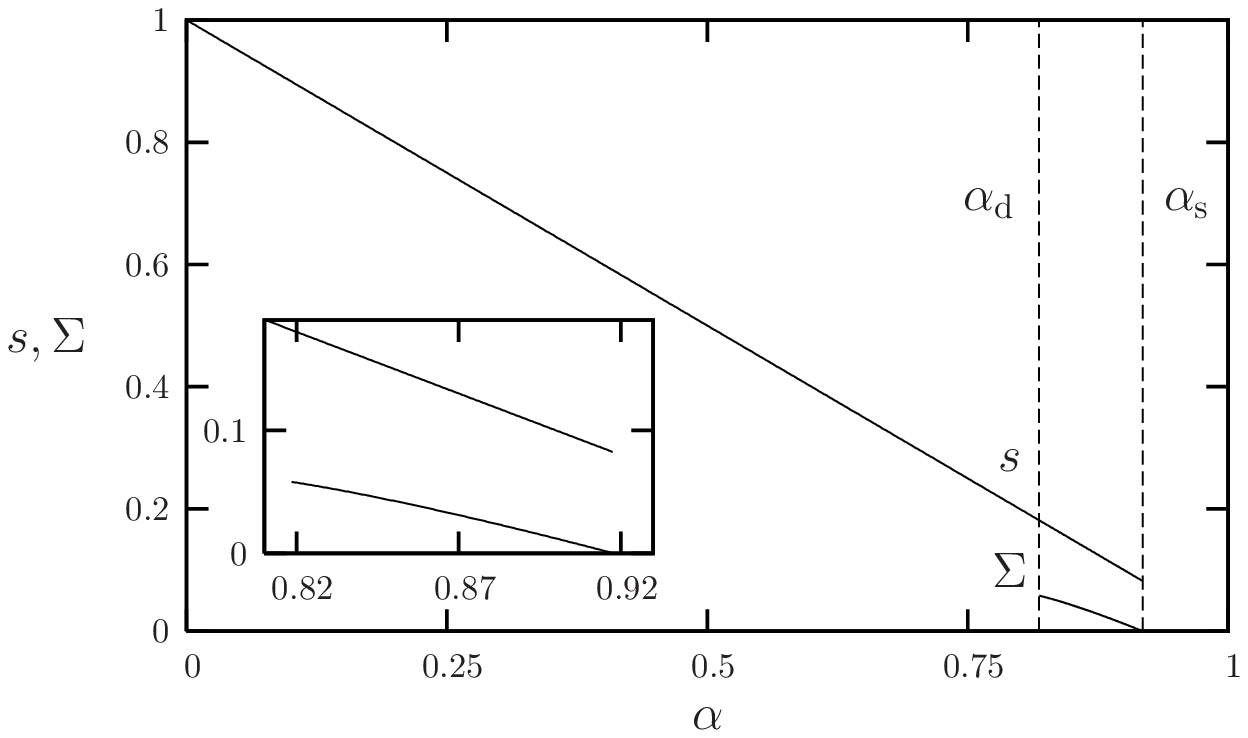}
\caption{Complexity and total entropy for 3-$\xorsat$, in units of $\ln
2$. The inset presents an enlargement of the regime 
$\alpha \in [\alpha_{\rm d},\alpha_{\rm s}]$.}
\label{fig_entropy_xorsat}
\end{figure}

It turns out that $F$ is satisfiable if and only if $F'$ is, and that
the number of solutions of these two formulas are related in an enlightening
way. It is clear that if the 2-core has no solution, there is no way to find
one for the full formula. Suppose on the contrary that an assignment of the 
variables in $V'$ that satisfy the equations of $F'$ has been
found, and let us show how to construct a solution of $F$
(and count in how many possible ways we can do this). Set ${\cal N}_0
=1$, and reintroduce step by step the removed equations, starting from
the last: in the $n$'th step of this new procedure we reintroduce the
clause which was removed at step $T_* - n$ of the leaf removal.  This
reintroduced clause has $d_n=|V_{T_*-n-1}| -|V_{T_*-n}| \ge 1 $
leaves; their configuration can be chosen in $2^{d_n -1}$ ways to
satisfy the reintroduced clause, irrespectively of the previous
choices, and we bookkeep this number of possible extensions 
by setting ${\cal N}_{n+1} = {\cal N}_n 2^{d_n
-1}$. Finally the total number of solutions of $F$ compatible with the
choice of the solution of $F'$ is obtained by adding the freedom of
the variables which appeared in no equations of $F$, ${\cal N}_{\rm
int} = {\cal N}_{T_*} 2^{N-|V_0|}$.  Let us underline that ${\cal
N}_{\rm int}$ is independent of the initial satisfying assignment of
the variables in $V'$, as appears clearly from the description of the
reconstruction algorithm; this property can be traced back to the linear
algebra structure of the problem. This suggests naturally the
decomposition of the total number of solutions of $F$ as the product
of the number of satisfying assignments of $V'$, call it ${\cal
N}_{\rm core}$, by the number of compatible full solutions ${\cal
N}_{\rm int}$. In terms of the associated entropy densities this
decomposition is additive
\begin{equation}
s = \Sigma + s_{\rm int} \ , \qquad \Sigma \equiv \frac{1}{N} \ln
{\cal N}_{\rm core} \ , \qquad s_{\rm int} \equiv \frac{1}{N} \ln
{\cal N}_{\rm int} \ ,
\end{equation}
where the quantity $\Sigma$ is the entropy density associated to the core of
the formula.  It is in fact much easier technically to compute the
statistical (with respect to the choice of the random formula $F$)
properties of $\Sigma$ and $s_{\rm int}$ once this decomposition has
been done (the fluctuations in the number of solutions is much smaller
once the non-core part of the formula has been removed). The outcome
of the computations~\cite{xor_1,xor_2} is the determination of the
threshold value $\alpha_{\rm s}$ for the appearance of a solution of
the 2-core $F'$ (and thus of the complete formula), along with
explicit formulas for the typical values of $\Sigma$ and $s$.  These two
quantities are plotted on Fig.~\ref{fig_entropy_xorsat}. The
satisfiability threshold corresponds to the cancellation of $\Sigma$:
the number of solutions of the core vanishes
continuously at $\alpha_{\rm s}$, while the total entropy remains finite 
because of the freedom of choice for the variables in the non-core part
of the formula.

On top of the simplification in the analytical determination of the
satisfiability threshold, this core decomposition of a formula 
unveils the change in the structure of the
set of solutions that occurs at $\alpha_{\rm d}$. Indeed, let us call
cluster all solutions of $F$ reconstructed from a common solution of
$F'$. Then one can show that this partition of the solution set of $F$
exhibits the properties exposed in Sec.~\ref{sec_clustering}, namely
that solutions are well-connected inside a cluster and separated from
one cluster to another.  The number of clusters is precisely equal to
the number of solutions of the core subformula, it thus undergoes a
drastic modification at $\alpha_{\rm d}$.  For smaller ratio of
constraints the core is typically empty, there is one single cluster
containing all solutions; when the threshold $\alpha_{\rm d}$ is
reached there appears an exponential numbers of clusters, the rate of
growth of this exponential being given by the complexity $\Sigma$.
Before considering the extension of this picture to random SAT problems, 
let us mention
that further studies of the geometry of the space of solutions of
random $\xorsat$ instances can be found in~\cite{MoSe,clus_xxorsat}.

\subsection{Phase transitions in random $\sat$}
\index{random problems!random k-SAT}

The possibility of a clustering transition in random $\sat$ problems was
first studied in~\cite{BiMoWe} by means of variational
approximations. Later developments allowed the computation of the complexity 
and, from the condition of its cancellation, the estimation of the 
satisfiability threshold $\alpha_{\rm s}$. This was first done
for $\k=3$ in~\cite{MeZe} and generalized for $\k\ge 4$
in~\cite{MeMeZe}, some of the values of $\alpha_{\rm s}$ thus computed
are reported in Tab.~\ref{tab_thresholds_sat}. A systematic expansion
of $\alpha_{\rm s}$ at large $\k$ was also performed in~\cite{MeMeZe}.
\begin{table}
\caption{Critical connectivities for the dynamical, condensation and
  satisfiability transitions for $\ksat$ random formulas.}
\begin{tabular}{| l | r | r | r |}
\hline & $\alpha_{\rm d}$~\cite{KrMoRiSeZd} & $\alpha_{\rm
 c}$~\cite{KrMoRiSeZd} & $\alpha_{\rm s}$\cite{MeMeZe} \\ \hline $\k=3$
 & $3.86$ & $3.86$ & $4.267$ \\ $\k=4$ & $9.38 $ & $9.547$ & $9.93$ \\
 $\k=5$ & $19.16$ & $20.80$ & $21.12$ \\ $\k=6$ & $36.53$ & $43.08$ &
 $43.4$ \\ \hline
\end{tabular}
\label{tab_thresholds_sat}
\end{table}

$\sat$ formulas do not share the linear algebra structure of $\xorsat$,
which makes the analysis of the clustering transition much more difficult,
and leads to a richer structure of the satisfiable phase 
$\alpha \le \alpha_{\rm s}$. The simple graph theoretic arguments are
not valid anymore, one cannot extract a core subformula from which the
partition of the solutions into clusters follows directly.
It is thus necessary to define them as a partition of the
solutions such that each cluster is well-connected and well-separated
from the other ones. A second complication arises: there is no reason
for the clusters to contain all the same number of solutions, as was
ensured by the linear structure of $\xorsat$. On the contrary, as was observed
in~\cite{BiMoWe} and in~\cite{MePaRi} for the similar random COL
problem, one faces a variety of clusters with various internal
entropies $s_{\rm int}$.
The complexity $\Sigma$ becomes a function of $s_{\rm int}$, in other words 
the number of clusters of internal entropy density $s_{\rm int}$ is
typically exponential, growing at the leading order like
$\exp[N\Sigma(s_{\rm int})]$. Drawing the consequences of these observations, a
refined picture of the satisfiable phase, and in particular the
existence of a new (so-called condensation) threshold 
$\alpha_{\rm c} \in [\alpha_{\rm d},\alpha_{\rm s}]$, was advocated
in~\cite{KrMoRiSeZd}. Let us briefly sketch some of these new features
and their relationship with the previous results of~\cite{MeZe,MeMeZe}.
Assuming the existence of a positive, concave, complexity function 
$\Sigma(s_{\rm int})$, continuously vanishing outside an interval of internal 
entropy densities $[s_-,s_+]$, the total entropy density is given by
\begin{equation}
s = \lim_{N \to \infty} \frac{1}{N} \ln \int_{s_-}^{s_+} ds_{\rm int} \ 
e^{N[\Sigma(s_{\rm int}) + s_{\rm int}]} \ .
\end{equation}
In the thermodynamic limit the integral can be evaluated with the Laplace 
method. Two qualitatively distinct situations can arise, whether the
integral is dominated by a critical point in the interior of the interval
$[s_-,s_+]$, or by the neighborhood of the upper limit $s_+$. In the former
case an overwhelming majority of the solutions are contained in an exponential
number of clusters, while in the latter the dominant contributions comes
from a sub-exponential number of clusters of internal entropy $s_+$, as
$\Sigma(s_+)=0$. The threshold $\alpha_{\rm c}$ separates the first regime
$[\alpha_{\rm d},\alpha_{\rm c}]$ where the relevant clusters are exponentially
numerous, from the second, condensated situation for 
$\alpha \in [\alpha_{\rm c},\alpha_{\rm s}]$ with a sub-exponential number of
dominant clusters\footnote{ This picture
is expected to hold for $\k \ge 4$; for $\k=3$, the dominant clusters
are expected to be of sub-exponential degeneracy in the
whole clustered phase, hence $\alpha_{\rm c}=\alpha_{\rm d}$ in this case.}.

The computations of~\cite{MeZe,MeMeZe} did not take into account the
distribution of the various internal entropies of the clusters,
which explains the discrepancy in the estimation of the
clustering threshold $\alpha_{\rm d}$ between \cite{MeZe,MeMeZe} and 
\cite{KrMoRiSeZd}. Let us however emphasize that this refinement of the
picture does not contradict the estimation of the satisfiability threshold
of~\cite{MeZe,MeMeZe}: the complexity computed in these works is 
$\Sigma_{\rm max}$, the maximal value of $\Sigma(s_{\rm int})$ reached at a 
local maximum with $\Sigma'(s)=0$, which indeed vanishes when the whole
complexity function disappears.

It is fair to say that the details of the picture proposed by
statistical mechanics studies have rapidly evolved in the last years,
and might still be improved. They rely indeed on self-consistent
assumptions which are rather tedious to check~\cite{MoPaRi}. Some
elements of the clustering scenario have however been established
rigorously in~\cite{clus_rig_xsat1,clus_rig_xsat2,clus_rig_Fede}, 
at least for large
enough $\k$.  In particular these works demonstrated, for some values
of $\k$ and $\alpha$ in the satisfiable regime, the existence of
forbidden intermediate Hamming distances between pairs of
configurations, which are either close (in the same cluster) or far
apart (in two distinct clusters).

Note finally that the consequences of such distributions of
clusters internal entropies were investigated on a toy model in~\cite{rcm},
and that yet another threshold $\alpha_{\rm f} > \alpha_{\rm d}$ 
for the appearance of frozen variables constrained to take the same
values in all solutions of a given cluster was investigated 
in~\cite{rearr_csp}.

\subsection{A glimpse at the computations}
\label{sec_computations}
\index{cavity method}
The statistical mechanics of disordered systems~\cite{Beyond} was first 
developed on so-called fully-connected models, where each variable appears
in a number of constraints which diverges in the thermodynamic limit. This
is for instance the case of the perceptron problem discussed in 
Sec.~\ref{sec_basic}. On the contrary, in a random $\k$-SAT instance a 
variable is typically involved in a finite number of clauses, one speaks in
this case of a diluted model. This finite connectivity is a
source of major technical complications. In particular the replica method,
alluded to in Sec.~\ref{bper} and applied to random $\k$-SAT 
in~\cite{MoZe,BiMoWe}, turns out to be rather cumbersome for diluted models 
in the presence of clustering~\cite{replica_diluted}. The cavity
formalism~\cite{cavity,cavity_T0,MeZe}, formally equivalent to the replica 
one, is more adapted to the diluted models.
In the following paragraphs
we shall try to give a few hints at the strategy underlying the
cavity computations, that might hopefully ease the reading of the original
literature.

The description of the random formula ensemble has two complementary
aspects: a global (thermodynamic) one, which amounts to the computation
of the typical energy and number of optimal configurations. A more
ambitious description will also provide geometrical information on
the organization of this set of optimal configurations inside the
$N$-dimensional hypercube. As discussed above these two aspects are in
fact interleaved, the clustering affecting both the thermodynamics (by
the decomposition of the entropy into the complexity and the internal
entropy) and the geometry of the configuration space. Let us for
simplicity concentrate on the $\alpha < \alpha_{\rm s}$ regime and consider a
satisfiable formula $F$. Both thermodynamic and geometric aspects can be
studied in terms of the uniform probability law over
the solutions of $F$:
\begin{equation}
\mu(\us) = \frac{1}{Z} \prod_{a=1}^M w_a(\us_a) \ ,
\label{eq_mu}
\end{equation}
where $Z$ is the number of solutions of $F$, the product runs over its
clauses, and $w_a$ is the indicator function of the event ``clause $a$
is satisfied by the assignment $\us$'' (in fact this depends only on
the configuration of the $\k$ variables involved in the clause $a$,
that we denote $\us_a$).  For instance the (information theoretic)
entropy of $\mu$ is equal to $\ln Z$, the log degeneracy of solutions, and
geometric properties can be studied by computing averages with respect to $\mu$
of well-chosen functions of $\us$.

\begin{figure}
\includegraphics[width=7cm]{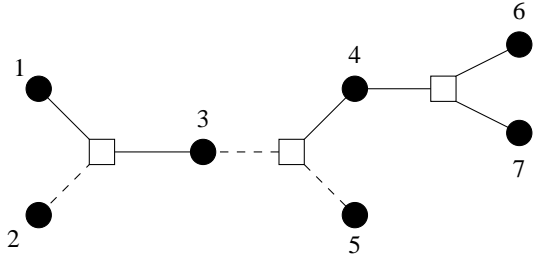}
\caption{The factor graph representation of a small 3-$\sat$ formula:
$(x_1 \vee \overline{x_2} \vee x_3 ) \wedge ( \overline{x_3} \vee x_4
\vee \overline{x_5} ) \wedge ( x_4 \vee x_6 \vee x_7)$.}
\label{fig_factor}
\end{figure}

A convenient representation of such a law is provided by factor
graphs~\cite{fgraphs}. These are bipartite graphs with two types of
vertices (see Fig.~\ref{fig_factor} for an illustration): one variable
node (filled circle) is associated to each of the $N$ Boolean
variables, while the clauses are represented by $M$ constraint nodes
(empty squares). By convention we use the indices $a,b,\dots$ for the 
constraint nodes, $i,j,\dots$ for the variables.
An edge is drawn between variable node $i$ and
constraint node $a$ if and only if $a$ depends on $i$. To
precise further by which value of $\s_i$ the clause $a$ gets
satisfied one can use two type of linestyles, solid and dashed on the
figure. A notation repeatedly used in the following is $\partial a$ 
(resp. $\partial i$) for the neighborhood of a constraint
(resp. variable) node, i.e. the set of adjacent variable
(resp. constraint) nodes. In this context $\setminus$ denotes the subtraction
from a set. We shall more precisely denote $\partial_+ i(a)$
(resp. $\partial_- i(a)$) the set of clauses in $\partial i \setminus a$ 
agreeing (resp. disagreeing) with $a$ on the satisfying value of
$\s_i$, and $\partial_\sigma i$ the set of clauses in
$\partial i$ which are satisfied by $\sigma_i=\sigma$.
This graphical representation naturally suggests a notion of distance
between variable nodes $i$ and $j$, 
defined as the minimal number of constraint nodes
crossed on a path of the factor graph linking nodes $i$ and $j$.

Suppose now that $F$ is drawn from the random ensemble. The corresponding 
random factor graph enjoys several interesting properties~\cite{random_graphs}.
The degree $|\partial i|$ of a randomly chosen variable $i$ is, in the 
thermodynamic limit, a Poisson random variable of average $\alpha \k$. If 
instead of a node one chooses randomly an edge $a-i$, the outdegree 
$|\partial i \setminus a|$ of $i$ has again a Poisson distribution with the
same parameter. Moreover the sign of the literals being chosen uniformly,
independently of the topology of the factor graph, the degrees 
$|\partial_+ i|$, $|\partial_- i|$, $|\partial_+ i(a)|$ and 
$|\partial_- i(a)|$ are Poisson random variables of parameter $\alpha \k /2$.
Another important feature of these random factor graphs is their local
tree-like character: if
the portion of the formula at graph distance smaller than $L$ of a
randomly chosen variable is exposed, the probability that this
subgraph is a tree goes to 1 if $L$ is kept fixed while the size $N$
goes to infinity.

\begin{figure}
\includegraphics[width=12cm]{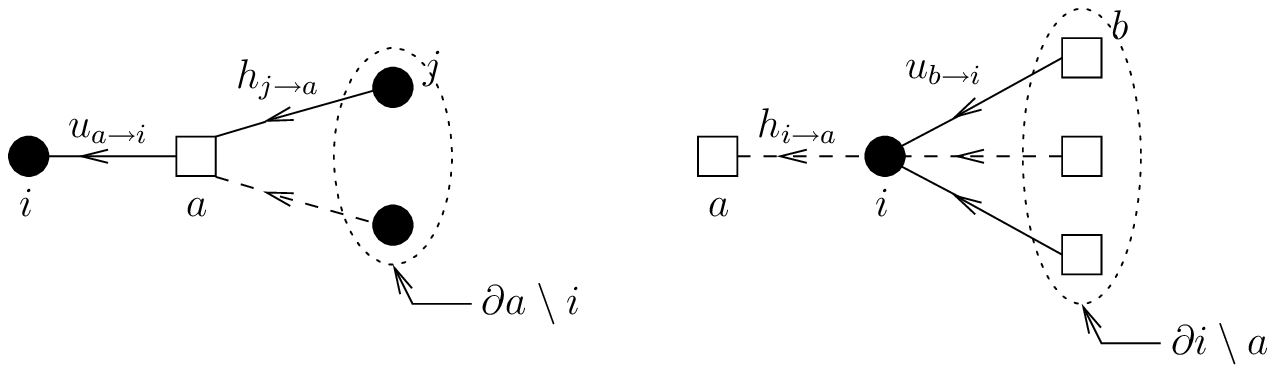}
\caption{A schematic representation of Eq.~(\ref{eq_recurs}).}
\label{fig_recurs}
\end{figure}

Let us for a second forget about the rest of the graph and consider a
finite formula whose factor graph is a tree, as is the case for the example
of Fig.~\ref{fig_factor}. The probability law $\mu$ of
Eq.~(\ref{eq_mu}) becomes in this case a rather simple object.  Tree
structures are indeed naturally amenable to a recursive (dynamic
programming) treatment, operating first on sub-trees which are then glued
together. More precisely, for each edge between a variable node $i$ and a 
constraint node $a$ one defines the amputated tree $F_{a \to i}$ 
(resp. $F_{i \to a}$) by removing all clauses in $\partial i$ apart from $a$
(resp. removing only $a$). These subtrees are associated to probability laws
$\mu_{a\to i}$ (resp. $\mu_{i \to a}$), defined as in Eq.~(\ref{eq_mu}) but
with a product running only on the clauses present in $F_{a \to i}$ 
(resp. $F_{i \to a}$). The marginal law of the root variable $i$ in these
amputated probability measures can be parametrized by a single real,
as $\sigma_i$ can take only two values (that, in the Ising spin convention, 
are $\pm 1$).
We thus define these fields, or messages, $h_{i \to a}$ and $u_{a \to i}$, by
\begin{equation}
\mu_{i \to a}(\sigma_i) = \frac{1- J_i^a \sigma_i \tanh h_{i \to
a}}{2} \ , \qquad \mu_{a \to i}(\sigma_i) = \frac{1- J_i^a \sigma_i
\tanh u_{a \to i}}{2} \ ,
\label{eq_cavitymarginals}
\end{equation}
where we recall that $\s_i=J_i^a$ is the value of the literal $i$
unsatisfying clause $a$.  A standard reasoning (see for
instance~\cite{BrMeZe}) allows to derive recursive equations 
(illustrated in Fig.~\ref{fig_recurs}) on these
messages, 
\begin{eqnarray}
h_{i \to a} &=& \sum_{b \in \partial_+ i(a)} u_{b \to i} - \sum_{b \in
\partial_- i(a)} u_{b \to i} \ ,  \label{eq_recurs} \\
u_{a \to i} &=& 
- \frac{1}{2} \ln \left(1 - \prod_{j \in \partial a \setminus i}
\frac{1-\tanh h_{j \to a}}{2} \right) \ .
\nonumber
\end{eqnarray}
Because the factor graph is a tree this set of equations has a unique 
solution which can be efficiently determined: one start from the leaves 
(degree 1 variable nodes) which obey the boundary condition $h_{i \to a} = 0$,
and progresses inwards the graph. The law $\mu$ can be completely
described from the values of the $h$'s and $u$'s solutions of these
equations for all edges of the graph. For instance the marginal
probability of $\sigma_i$ can be written as
\begin{equation}
\mu(\sigma_i) = \frac{1 + \sigma_i \tanh h_i }{2} \ , \qquad h_i =
\sum_{a \in \partial_+ i} u_{a \to i} - \sum_{a \in \partial_- i} u_{a
\to i} \ .
\label{eq_mui}
\end{equation}
In addition the entropy $s$ of solutions of
such a tree formula, can be computed from the values of the
messages $h$ and $u$ \cite{BrMeZe}.
\comment{\begin{equation}
N s = \ln Z = - \sum_{a-i} \ln z_{a-i}(h_{i \to a},u_{a \to i}) +
\sum_a \ln z_a(\{ h_{i \to a} \}_{i \in \partial a} ) + \sum_i \ln
z_i(\{ u_{a \to i}\}_{a \in \partial i} ) \ ,
\label{eq_Bethe}
\end{equation}
where the three sum runs respectively over the edges, constraint nodes and
variable nodes of the factor graph, and the various functions $z$'s having 
simple expressions~\cite{BrMeZe} that for simplicity we do not give 
explicitly here.}

We shall come back to the equations (\ref{eq_recurs}), and justify the
denomination messages, in Sec.~\ref{sec_mp}; these can be interpreted as
the Belief Propagation~\cite{fgraphs,Yedidia,Yedidia2} heuristic equations for
loopy factor graphs.

The factor graph of random formulas is only locally tree-like;
the simple computation sketched above has thus to be amended in order 
to take into account the effect of the distant, loopy part of the formula. 
Let us call $F_L$ the factor graph made of variable nodes at graph distance
smaller than or equal to $L$ from an arbitrarily chosen variable node $i$ in a
large random formula $F$, and $B_L$ the variable nodes at distance
exactly $L$ from $i$. Without loss of generality in the thermodynamic limit,
we can assume that $F_L$ is a tree. The cavity method amounts to an hypothesis 
on the effect of the distant part of the factor graph, $F \setminus F_L$, 
i.e. on the boundary condition it induces on $F_L$. 
In its simplest (so called replica symmetric) version, that is believed to 
correctly describe the unclustered situation for $\alpha \le \alpha_{\rm d}$, 
$F \setminus F_L$ is replaced, for each variable node $j$ in the boundary
$B_L$, by a fictitious constraint node which sends a bias 
$u_{{\rm ext} \to j}$. In other words the boundary condition is factorized
on the various nodes of $B_L$; such a simple description
is expected to be correct for $\alpha \le \alpha_{\rm d}$ because,
in the amputated factor graph $F \setminus F_L$, the distance between the 
variables of $B_L$ is typically large (of order $\ln N$), and
these variables should thus be weakly correlated. These external biases
are then turned into random variables to take into account the randomness
in the construction of the factor graphs, and Eq.~(\ref{eq_recurs})
acquires a distributional meaning. The messages $h$ (resp. $u$) are
supposed to be i.i.d. random variables drawn from a common distribution, the
degrees $\partial_\pm i(a)$ being two independent Poisson random variables
of parameter $\alpha \k /2$. These distributional equations can be 
numerically solved by a population dynamics algorithm~\cite{cavity}, also
known as a particle representation in the statistics litterature.
The typical entropy density is then computed by averaging $s$
over these distributions of $h$ and $u$.

This description fails in the presence of clustering, which induces
correlations between the variable nodes of $B_L$ in the amputated factor 
graph $F \setminus F_L$. To take these correlations
into account a refined version of the cavity method (termed one step
of replica symmetry breaking, in short 1RSB) has been developed. It
relies on the hypothesis that the partition of the solution space into
clusters $\gamma$ has nice decorrelation properties: once decomposed
onto this partition, $\mu$ restricted to a cluster $\gamma$ behaves
essentially as in the unclustered phase (it is a pure state in
statistical mechanics jargon). Each directed edge $a \to i$ should thus
bear a family of messages $u_{a \to i}^\gamma$, one for each cluster,
or alternatively a distribution $Q_{a \to i}(u)$ of the messages with respect 
to the choice of $\gamma$. The equations (\ref{eq_recurs}) are thus
promoted to recursions between distributions $P_{i \to a}(h)$, 
$Q_{a \to i}(u)$, which depends on a real $m$ known as the Parisi
breaking parameter. Its role is to select the size of the investigated 
clusters, i.e. the number of solutions they contain. The computation
of the typical entropy density  is indeed replaced
by a more detailed thermodynamic potential,
\begin{equation}
\Phi(m) = \frac{1}{N} \ln \sum_\gamma Z_\gamma^m = 
\frac{1}{N} \ln \int_{s_-}^{s_+} ds_{\rm int} \ 
e^{N [\Sigma(s_{\rm int}) + m s_{\rm int} ]} \ .
\label{eq_Phi}
\end{equation}
In this formula $Z_\gamma$ denotes the number of solutions inside a cluster 
$\gamma$, and we used the hypothesis that at the leading order the number
of clusters with internal entropy density $s_{\rm int}$ is given by
$\exp[N \Sigma(s_{\rm int})]$. The complexity function $\Sigma(s_{\rm int})$
can thus be obtained from $\Phi(m)$ by an inverse Legendre transform. 
For generic values of $m$ this approach is computationally very demanding;
following the same steps as in the replica symmetric version of the
cavity method one faces a distribution (with respect to the topology of
the factor graph) of distributions (with respect to the choice of the
clusters) of messages. Simplifications however arise for $m=1$ and 
$m=0$~\cite{KrMoRiSeZd}; the latter case corresponds in fact to the
original Survey Propagation approach of~\cite{MeZe}. As appears clearly
in Eq.~(\ref{eq_Phi}), for this value of $m$ all clusters are treated on an
equal footing and the dominant contribution comes from the most numerous
clusters, independently of their sizes. Moreover, as we further explain
in Sec.~\ref{sec_mp}, the structure of the equations can be greatly simplified
in this case, the distribution over the cluster of fields being parametrized
by a single number.

\subsection{Finite Size Scaling results}
\label{sec_review_FSS}
\index{finite size scaling}

As we explained in Sec.~\ref{sec_gendef} the threshold phenomenon can
be more precisely described by finite size scaling relations. Let us
mention some FSS results about the transitions we just discussed.

For random 2-$\sat$, where the satisfiability property is
known~\cite{transition_k2} to exhibit a sharp threshold at
$\alpha_{\rm s}=1$, the width of the transition window has been
determined in~\cite{FSS_k2}.  The range of $\alpha$ where the
probability of satisfaction drops significantly is of order
$N^{-1/3}$, i.e. the exponent $\nu$ is equal to $3$, as for the random
graph percolation. This similarity is not surprising, the proof
of~\cite{FSS_k2} relies indeed on a mapping of 2-$\sat$ formulas onto
random (directed) graphs.

The clustering transition for $\xorsat$ was first conjectured
in~\cite{FSS_codes} (in the related context of error-correcting codes)
then proved in~\cite{FSS_cores} to be described by
\begin{equation}
P(N,M=N(\alpha_{\rm d}+N^{-1/2}\lambda + N^{-2/3} \delta) ) = {\cal
F}(\lambda) + O(N^{-5/26}) \ ,
\end{equation}
where $\delta$ is a subleading shift correction that has been explicitly
computed, and the scaling function $\cal F$ is, upto a multiplicative
factor on $\lambda$, the same error function as in Eq.~(\ref{percep_fss}).

A general result has been proved in~\cite{FSS_Wilson} on the width of
transition windows. Under rather unrestrictive conditions one can show
that $\nu \ge 2$: the transitions cannot be arbitrarily sharp. Roughly
speaking the bound is valid when a finite fraction of the clauses are
not decisive for the property of the formulas studied, for instance
clauses containing a leaf variable are not relevant for the
satisfiability of a formula. The number of these irrelevant clauses is
of order $N$ and has thus natural fluctuations of order $\sqrt{N}$;
these fluctuations blur the transition window which cannot be sharper
than $N^{-1/2}$.

Several studies (see for
instance~\cite{FSS_KiSe,FSS_Moetal,xor_replica}) have attempted to
determine the transition window from numeric evaluations of the
probability $P(N,\alpha)$, for instance for the satisfiability
threshold of random 3-$\sat$~\cite{FSS_KiSe,FSS_Moetal} and
$\xorsat$~\cite{xor_replica}.  These studies are necessarily confined to
small formula sizes, as the typical computation cost of complete
algorithms grows exponentially around the transition. In consequence
the asymptotic regime of the transition window, $N^{-1/\nu}$, is often
hidden by subleading corrections which are difficult to evaluate, and
in~\cite{FSS_KiSe,FSS_Moetal} the reported values of $\nu$ were found
to be in contradiction with the latter derived rigorous bound. This is
not an isolated case, numerical studies are often plagued by
uncontrolled finite-size effects, as for instance in the bootstrap
percolation~\cite{bootstrap}, a variation of the classical percolation
problem.

\section{Local search algorithms}  %
\label{sec_localsearch}

The following of this review will be devoted to the study of various
solving algorithms for $\sat$ formulas. Algorithms are, to some extent,
similar to dynamical processes studied in statistical physics. In this
context the focus is however mainly on stochastic processes that
respect detailed balance with respect to the Gibbs-Boltzmann 
measure~\cite{Leticia}, a condition which is rarely respected by solving
algorithms. Physics inspired techniques can yet be useful, and will
emerge in three different ways. The random walk algorithms considered in this 
Section are stochastic processes in the space of configurations (not fulfilling
the detailed balance condition), moving by small steps where one or a few 
variables are modified. 
Out-of-equilibrium physics (and in particular growth processes) provide
an interesting view on classical complete algorithms (DPLL), as shown in 
Sec.~\ref{sec_DPLL}. Finally, the picture of the satisfiable phase put
forward in Sec.~\ref{sec:phase_transitions} underlies the message-passing
procedures discussed in Sec.~\ref{sec_mp}.

\subsection{Pure random walk sat, definition and results valid for all 
instances}

Papadimitriou~\cite{Papadimitriou} proposed the following algorithm,
called Pure Random Walk Sat (PRW$\sat$) in the following, to solve
$\ksat$ formulas:

\begin{enumerate}
\item Choose an initial assignment $\us(0)$ uniformly at random and set
$T=0$.
\item If $\us(T)$ is a solution of the formula (i.e. $E(\us(T))=0$),
output {\scshape solution} and stop. If $T=T_{\rm max}$, a threshold
fixed beforehand, output {\scshape undetermined} and stop.
\item Otherwise, pick uniformly at random a clause among those that
are $\unsat$ in $\us(T)$; pick uniformly at random one of the $\k$
variables of this clause and flip it (reverse its status from True to
False and vice-versa) to define the next assignment $\us(T+1)$; set $T
\to T+1$ and go back to step 2.
\end{enumerate}

This defines a stochastic process $\us(T)$, a biased random walk in
the space of configurations. The modification $\us(T) \to \us(T+1)$ in
step 3 makes the selected clause satisfied; however the flip of a
variable $i$ can turn previously satisfied clauses into unsatisfied
ones (those which were satisfied solely by $i$ in $\us(T)$).

This algorithm is not complete: if it outputs a solution one is
certain that the formula was satisfiable (and the current
configuration provides a certificate of it), but if no solution has
been found within the $T_{\rm max}$ allowed steps one cannot be sure
that the formula was unsatisfiable. There are however two rigorous
results which makes it a probabilistically almost complete
algorithm~\cite{random_algo}.

For $\k=2$, it was shown in~\cite{Papadimitriou} that PRW$\sat$ finds a
solution in a time of order $O(N^2)$ with high probability for all
satisfiable instances. Hence, one is almost certain that the formula
was unsatisfiable if the output of the algorithm is {\scshape
undetermined} after $T_{\rm max} = O(N^2)$ steps.

Sch\"oning~\cite{schoning} proposed the following variation for $\k=3$.
If the algorithm fails to find a solution before $T_{\rm max} = 3N$ steps,
instead of stopping and printing {\scshape undetermined}, it restarts
from step 1, with a new random initial condition $\us(0)$. Sch\"oning
proved that if after $R$ restarts no solution has been found, then the
probability that the instance is satisfiable is upper-bounded by $\exp
[-R\times(3/4)^N]$ (asymptotically in $N$). This means that a
computational cost of order $(4/3)^N$ allows to reduce the probability
of error of the algorithm to arbitrary small values. Note that if the time
scaling of this bound is exponential, it is also exponentially smaller than 
the $2^N$ cost of an exhaustive enumeration. Improvements on the factor
$4/3$ are reported in~\cite{schoning2}.

\subsection{Typical behavior on random $\ksat$ instances}
\label{secrwsat2}

\begin{figure}
\includegraphics[width=8.5cm]{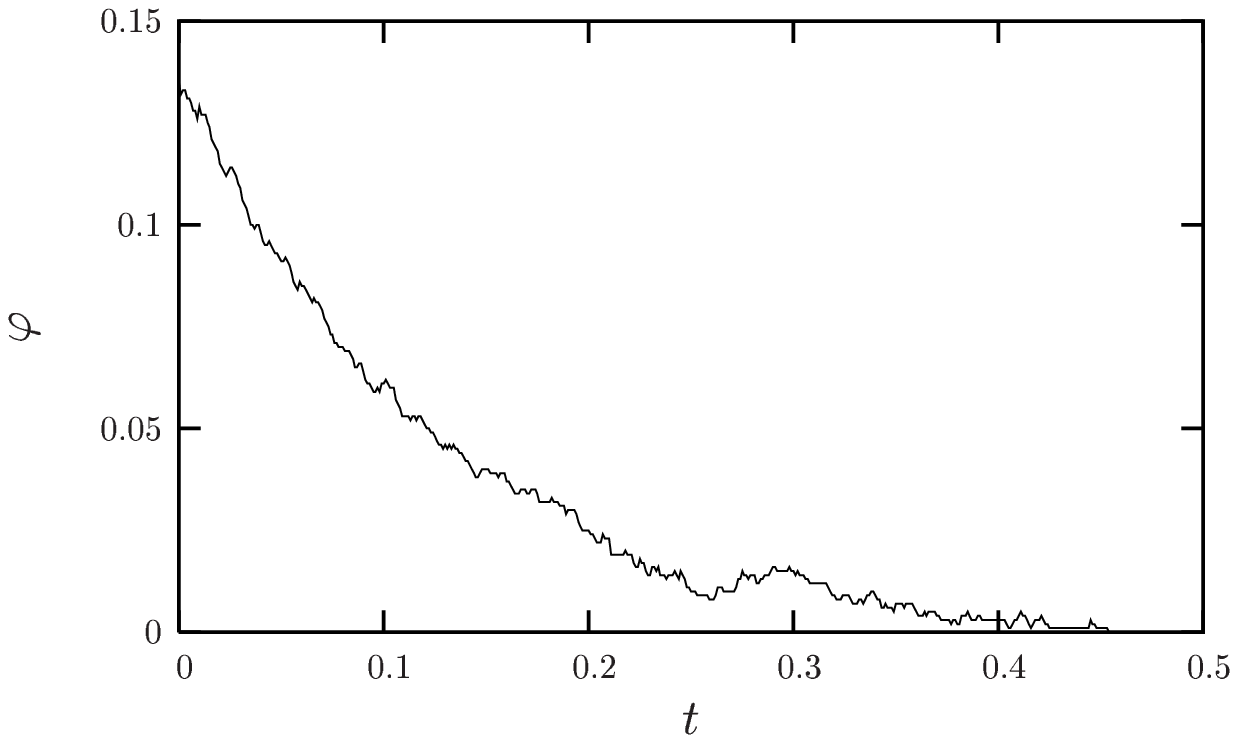}

\includegraphics[width=8.5cm]{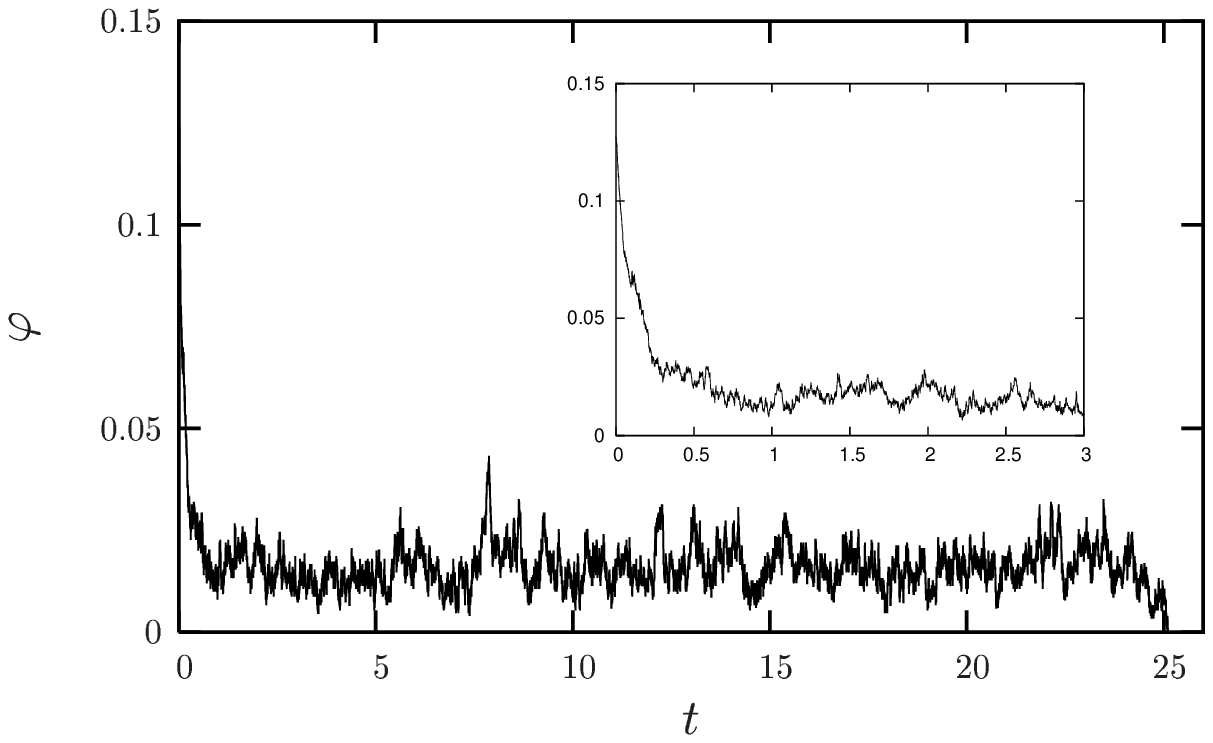}
\caption{Fraction of unsatisfied constraints $\varphi=E/M$ in function
  of reduced time $t=T/M$ during the execution of PRW$\sat$ on random
  3-$\sat$ formulas with $N=500$ variables. Top: $\alpha=2$, Bottom:
  $\alpha=3$.}
\label{fig_prwsat_1}
\end{figure}

The results quoted above are true for any $\ksat$ instance. An
interesting phenomenology arises when one applies the PRW$\sat$ algorithm
to instances drawn from the random $\ksat$
ensemble~\cite{ws1,ws2}. Figure~\ref{fig_prwsat_1} displays the
temporal evolution of the number of unsatisfied clauses during the
execution of the algorithm, for two random 3-$\sat$ instances of
constraint ratio $\alpha=2$ and $3$. The two curves are very
different: at low values of $\alpha$ the energy decays rather fast
towards 0, until a point where the algorithm finds a solution and
stops. On the other hand, for larger values of $\alpha$, the energy
first decays towards a strictly positive value, around which it
fluctuates for a long time, until a large fluctuation reaches 0,
signaling the discovery of a solution. A more detailed study with
formulas of increasing sizes reveals that a threshold value
$\alpha_{\rm rw} \approx 2.7$ (for $\k=3$) sharply separates this two
dynamical regimes. In fact the fraction of unsatisfied clauses
$\varphi=E/M$, expressed in terms of the reduced time $t=T/M$,
concentrates in the thermodynamic limit around a deterministic
function $\varphi(t)$. For $\alpha < \alpha_{\rm rw}$ the function
$\varphi(t)$ reaches 0 at a finite value $t_{\rm sol}(\alpha,\k)$,
which means that the algorithm finds a solution in a linear number of
steps, typically close to $N t_{\rm sol}(\alpha,\k)$. 
On the contrary for $\alpha
> \alpha_{\rm rw}$ the reduced energy $\varphi(t)$ reaches a positive
value $\varphi_{\rm as}(\alpha,\k)$ as $t \to \infty$; a solution, if
any, can be found only through large fluctuations of the energy which
occur on a time scale exponentially large in $N$. This is an
example of a metastability phenomenon, found in several other
stochastic processes, for instance the contact process~\cite{cp}. When
the threshold $\alpha_{\rm rw}$ is reached from below the solving time
$t_{\rm sol}(\alpha,\k)$ diverges, while the height of the plateau
$\varphi_{\rm as}(\alpha,\k)$ vanishes when $\alpha_{\rm rw}$ is
approached from above.

\begin{figure}
\includegraphics[width=8.5cm]{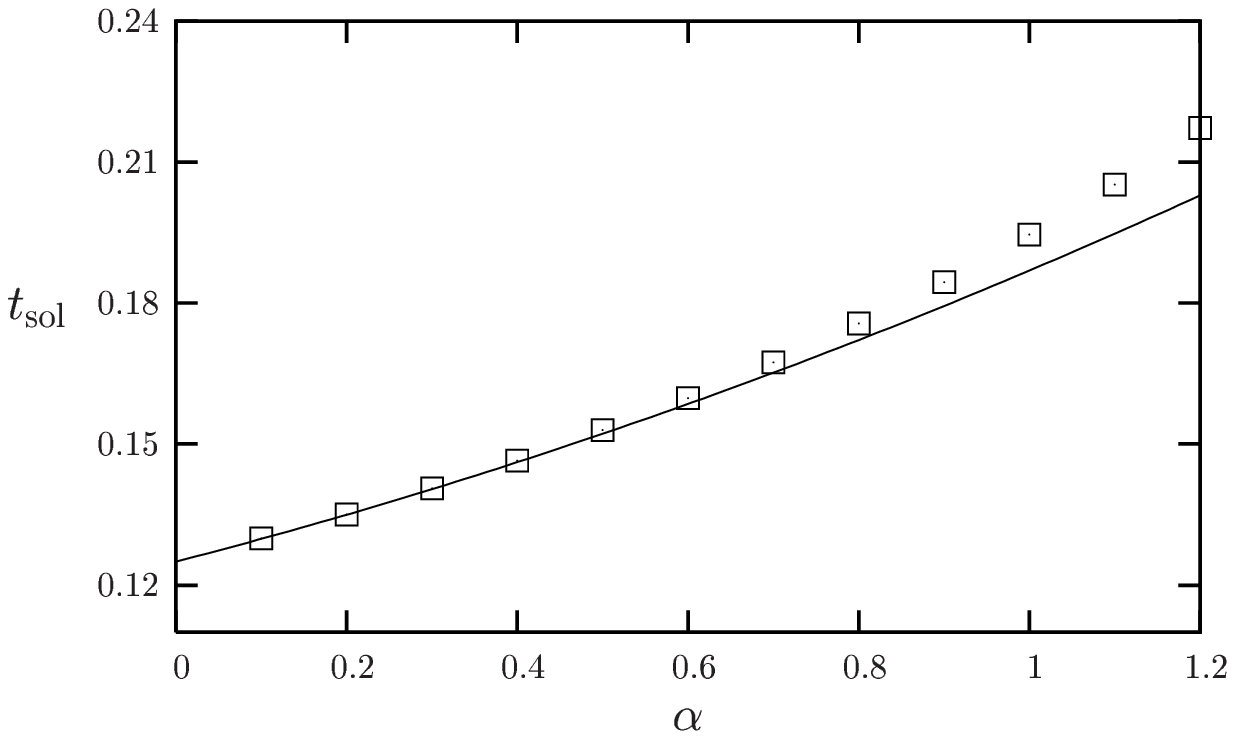}

\includegraphics[width=8.5cm]{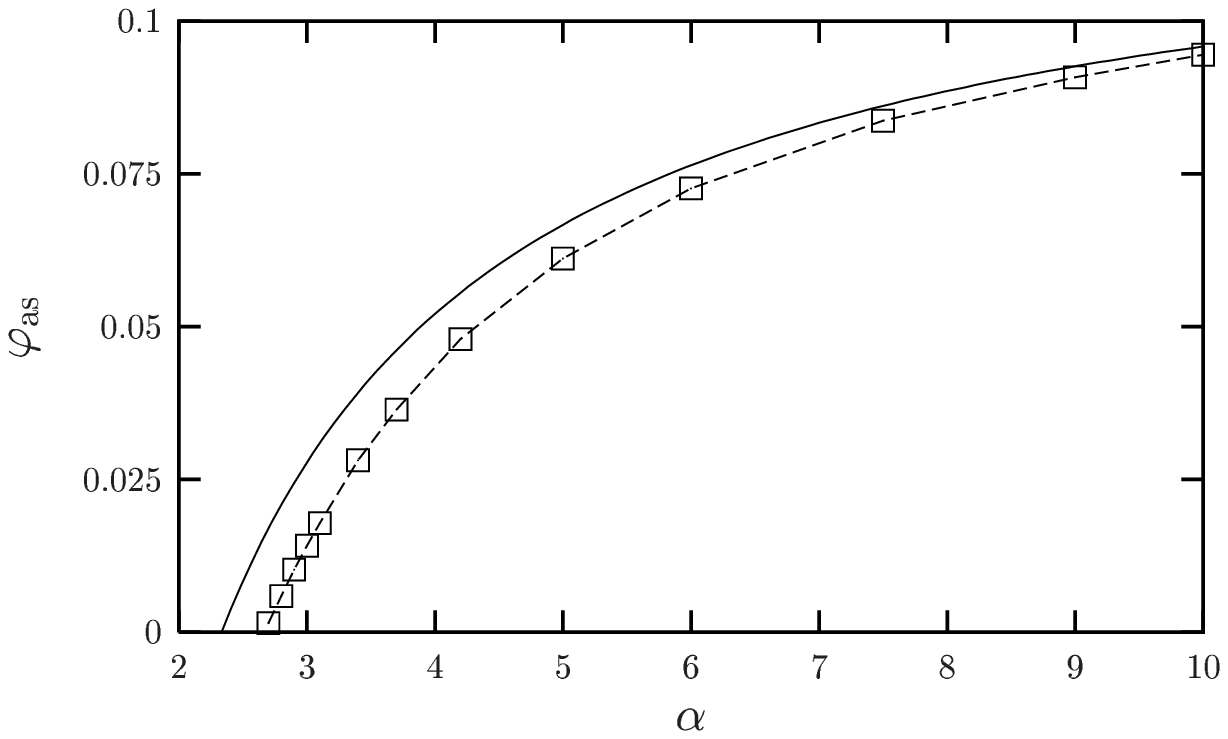}
\caption{Top: linear solving time $t_{\rm sol}(\alpha,3)$ for random
  3-$\sat$ formulas in function of $\alpha$; symbols correspond to
  numerical simulations, solid line to the second order expansion in
  $\alpha$ obtained in~\cite{ws1}. Bottom: fraction of unsatisfied
  constraints reached at large time for $\alpha > \alpha_{\rm rw}$ for
  random 3-$\sat$ formulas; symbols correspond to numerical
  simulations, solid line to the approximate analytical computations
  of~\cite{ws1,ws2}.}
\label{fig_prwsat_2}
\end{figure}

In~\cite{ws1,ws2} various statistical mechanics inspired techniques 
have been applied to study
analytically this phenomenology, some results are presented in
Figure~\ref{fig_prwsat_2}.  The low $\alpha$ regime can be tackled by
a systematic expansion of $t_{\rm sol}(\alpha,\k)$ in powers of
$\alpha$. The first three terms of these series have been computed,
and are shown on the left panel to be in good agreement with the
numerical simulations.

Another approach was followed to characterize the transition
$\alpha_{\rm rw}$, and to compute (approximations of) the asymptotic
fraction of unsatisfied clauses $\varphi_{\rm as}$ and the intensity
of the fluctuations around it. The idea is to project the Markovian
evolution of the configuration $\us(T)$ on a simpler observable, the
energy $E(T)$. Obviously the Markovian property is lost in this
transformation, and the dynamics of $E(T)$ is much more complex. One
can however approximate it by assuming that all configurations of the
same energy $E(T)$ are equiprobable at a given step of execution of
the algorithm. This rough approximation of the evolution of $E(T)$ is
found to concentrate around its mean value in the thermodynamic limit,
as was constated numerically for the original process. Standard
techniques allow to compute this average approximated evolution, which
exhibits the threshold behavior explained above at a value $\alpha =
(2^\k -1)/\k$ which is, for $\k=3$, slightly lower than the threshold
$\alpha_{\rm rw}$. The right panel of Fig.~\ref{fig_prwsat_2}
confronts the results of this approximation with the numerical
simulations; given the roughness of the hypothesis the agreement is
rather satisfying, and is expected to improve for larger values of
$\k$.

The rigorous results on the behavior of PRW$\sat$ on random instances
are very few. Let us mention in particular~\cite{alek}, which proved
that the solving time for random 3-$\sat$ formulas is typically
polynomial up to $\alpha=1.63$, a result in agreement yet weaker than
the numerical results presented here.

\subsection{More performant variants of the algorithm}

The threshold $\alpha_{\rm rw}$ for linear time solving of random
instances by PRW$\sat$ was found above to be much smaller than the
satisfiability threshold $\alpha_{\rm s}$. It must however be
emphasized that PRW$\sat$ is only the simplest example of a large family
of local search algorithms, see for 
instance~\cite{skc,skc2,rrt,asat,circumspect}. They
all share the same structure: a solution is searched through a random
walk in the space of configurations, one variable being modified at
each step. The choice of the flipped variable is made according to
various heuristics; the goal is to find a compromise between the
greediness of the walk which seeks to minimize locally the energy of
the current assignment, and the necessity to allow for moves
increasing the energy in order to avoid the trapping in local minima
of the energy function. A frequently encountered ingredient of the 
heuristics, which is of a greedy nature, is the focusing: the flipped 
variable necessarily belongs to at least one unsatisfied clause before
the flip, which thus becomes satisfied after the move. Moreover, instead
of choosing randomly one of the $\k$ variables of the unsatisfied clause,
one can consider for each of them the effect of the flip, and avoid variables
which, once flipped, will turn satisfied clauses into unsatisfied 
ones~\cite{skc,skc2}. Another way to implement the greediness~\cite{rrt}
consists in bookkeeping the lowest energy found so far during the walk, and
forbids flips which will raise the energy of the current assignment above
the registered record plus a tolerance threshold. These demanding requirements
have to be balanced with noisy, random steps, allowing to escape traps which
are only locally minima of the objective function.

These more elaborated heuristics are very numerous, and depend on parameters
that are finely tuned to achieve the best performances, hence an exhaustive
comparison is out of the scope of this review. Let us only mention that some
of these heuristics are reported in~\cite{rrt,asat} to efficiently find 
solutions of large (up to $N=10^6$) random formulas of 3-$\sat$ at ratio 
$\alpha$ 
very close to the satisfiability threshold, i.e. for $\alpha \lesssim 4.21$.

\index{local search!WalkSAT}

\section{Decimation based algorithms} %
\label{sec_decimation}

The algorithms studied in the remaining of the review are of a very different
nature compared to the local search procedures described above. Given an
initial formula $F$ whose satisfiability has to be decided, they proceed by
assigning sequentially the value of some of the variables. The formula can be
simplified under such a partial assignment: clauses which are satisfied by at
least one of their literal can be removed, while literals unsatisfying a
clause are discarded from the clause. 
It is instructive to consider the following
thought experiment: 
suppose one can consult an oracle who, given a formula, is
able to compute the marginal probability of the variables, in the
uniform probability measure over the optimal assignments of the formula. With
the help of such an oracle it would be possible to sample uniformly the
optimal assignments of $F$, by computing these marginals, setting one
unassigned variable according to its marginal, and then proceed in the same
way with the simplified formula. A slightly less ambitious, yet still
unrealistic, task is to find one optimal configuration (not necessarily
uniformly distributed) of $F$; this can be performed if the oracle is able to
reveal, for each formula he is questioned about, which of the unassigned
variables take the same value in all optimal assignments, and what is this
value. Then it is enough to avoid setting incorrectly such a constrained
variable to obtain at the end an optimal assignment.

Of course such procedures are not meant as practical algorithms; instead of
these fictitious oracles one has to resort to simplified evidences gathered
from the current formula to guide the choice of the variable to assign.
In Sec.~\ref{secuc} we consider algorithms exploiting basic information on the
number of occurrences of each variable, and their behavior in the satisfiable
regime of random $\sat$ formulas. They are turned into complete algorithms by
allowing for backtracking the heuristic choices, as explained
in~\ref{sec_DPLL}. Finally in Sec.~\ref{sec_mp} we shall use more refined
message-passing sub-procedures to provide the information used in the
assignment steps.

\subsection{Heuristic search: the success-to-failure transition}
\label{secuc}
\index{unit propagation}

The first algorithm we consider was introduced and analyzed by Franco
and his collaborators \cite{Ch90_2,Ch90_1}.

\begin{enumerate}
\item If a formula contains a \emph{unit clause} i.e. a clause with a
          single variable, this clause is satisfied through an
          appropriate assignment of its unique variable (propagation); 
If the formula contains no {\em unit-clause} a variable and its
  truth value are chosen according to some heuristic rule (free choice).
Note that the unit clause propagation corresponds to the obvious answer an
          oracle would provide on such a formula.
        \item Then the clauses in which the assigned variable appears are
          simplified: satisfied clauses are removed, the other ones
          are reduced.
\item Resume from step 1.
\end{enumerate}

The procedure will end if one of two conditions is verified:

\begin{enumerate}
        \item The formula is completely empty (all clauses have been
        removed), and a solution has been found ({\sc success}).
        \item A contradiction is generated from the presence of two
        opposite unit clauses. The algorithm halts. We do not know if
        a solution exists and has not been found or if there is no
        solution ({\sc failure}).
\end{enumerate}

The simplest example of heuristic is called
Unit Clause (UC) and consists in choosing a variable uniformly at
random among those that are not yet set, and assigning it to {\scshape
true} or {\scshape false} uniformly at random. More sophisticated
heuristics can take into account the number of occurrences of each
variable and of its negation, the length of the clauses in which each
variable appears, or they can set more than one variable at a
time. For example, in the Generalized Unit Clause (GUC), the variable
is always chosen among those appearing in the shortest clauses.

Numerical experiments and theory show that the results of this procedure 
applied to random
$\ksat$ formulas with ratios $\alpha$ and size $N$ can be classified 
in two regimes:
\begin{itemize}
\item At low ratio $\alpha < \alpha_H$ the search procedure finds 
a solution with positive probability (over the formulas and the random
choices of the algorithm) when $N\to \infty$.
\item At high ratio $\alpha > \alpha_H$ the probability of finding a
  solution vanishes when $N\to\infty$. Notice that $\alpha_H <
  \alpha_{\rm s}$: solutions do exist in the range $[\alpha_H,\alpha_{\rm s}]$ 
but are not found by this heuristic.
\end{itemize}

The above algorithm \emph{modifies} the formula as it proceeds; during
the execution of the algorithm the current formula will contain
clauses of length 2 and 3 (we specialize here to $\k=3$-$\sat$ for the sake of
simplicity but higher values of $\k$ can be considered). 
The sub-formulas generated by the search procedure
maintain their statistical uniformity (conditioned on the number of clauses
of length 2 and 3). 
Franco and collaborators used this fact to write down differential
equations for the evolution of the densities 
of 2- and 3-clauses as a function of the fraction $t$ of eliminated
variables. We do not reproduce those equations here, see \cite{Achltcs}
for a pedagogical review. Based on this analysis Frieze and Suen
\cite{Fr96} were
able to calculate, in the limit of infinite size, the  probability
of successful search. The outcome for the UC heuristic is 
\begin{eqnarray}
        \mathcal P_{\rm success}^\mathrm{(UC)}(\alpha) = \exp
        \left\{ -\frac 1 {4 \sqrt{8 / 3 \alpha - 1}} \arctan \left[
        \frac 1 {\sqrt{8 / 3 \alpha - 1}} \right] - \frac 3 {16}
        \alpha \right\}
        \label{p_success_UC}
\end{eqnarray}
when $\alpha < \frac 83$, and ${\cal P}=0$ for larger ratios. The
probability ${\cal P}_\mathrm{success}$ is, as expected, a decreasing 
function of $\alpha$; it vanishes in $\alpha _H=\frac 83$.
A similar calculation shows that $\alpha _H \simeq 3.003$
for the GUC heuristic \cite{Fr96}.

\begin{figure}
\includegraphics[width=12cm]{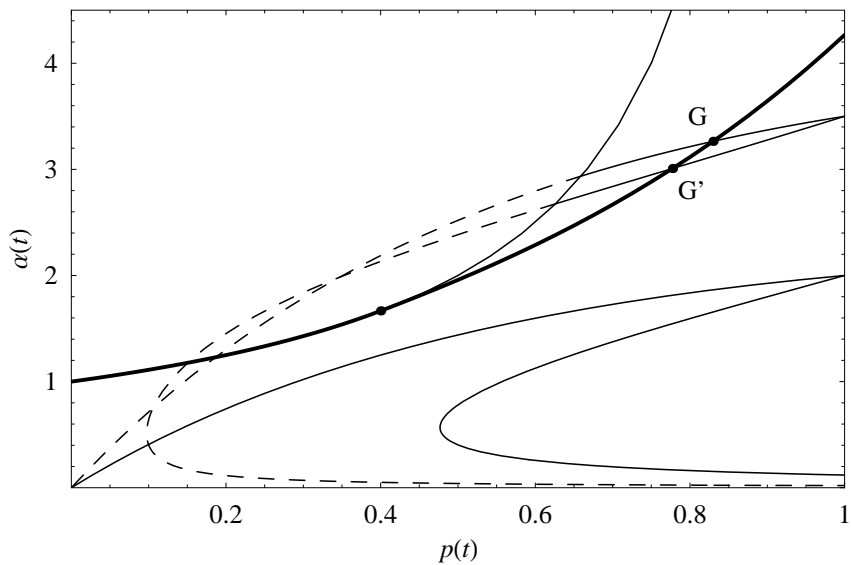}
\caption{Trajectories generated by heuristic search acting on 3-$\sat$ for 
$\alpha=2$ and $\alpha=3.5$. For all heuristics, the starting point is on the 
$p=1$ axis, with the initial value of $\alpha$ as ordinate. The curves that 
end at the origin correspond to UC, those ending on the $p=1$ axis correspond 
to GUC. The thick line represents the satisfiability threshold: the part on 
the left of the critical point $(2/5,5/3)$ is exact and coincides with the 
contradiction line, 
where contradictions are generated with high probability, of equation 
$\alpha = 1/ (1-p)$, and which is plotted for larger values of $p$ as well; 
the part on the right of the critical point is only a sketch. When the 
trajectories hit the satisfiability threshold, at points G for UC and G' for 
GUC, they enter a region in which massive backtracking takes place, and the 
trajectory represents the evolution \emph{prior} to backtracking. The 
dashed part of the curves is ``unphysical'', i.e. the trajectories stop when 
the contradiction curve is reached.
}
\label{fig_trajectories}
\end{figure}

Franco et al's analysis can be recast in the
following terms. Under the operation of the algorithm the original
3-$\sat$ formula is turned into a mixed $2+p$-$\sat$ formula where $p$
denotes the fraction of the clauses with 3 variables: there are
$N \alpha \cdot (1-p)$ 2-clauses and $N \alpha p$ 3-clauses.
As we mentioned earlier the simplicity of the heuristics maintains
a statistical uniformity over the formulas with a given value of $\alpha$
and $p$. This constatation motivated the study of the random $2+p$-$\sat$
ensemble by statistical mechanics methods~\cite{FSS_Moetal,BiMoWe}, some
of the results being later confirmed by the rigorous analysis 
of~\cite{2pp_rigorous}. At the heuristic level one expects the existence of
a $p$ dependent satisfiability threshold $\alpha_{\rm s}(p)$, interpolating
between the 2-$\sat$ known threshold, $\alpha_{\rm s}(p=0)=1$, and the 
conjectured 3-$\sat$ case, $\alpha_{\rm s}(p=1)\approx 4.267$. The upperbound
$\alpha_{\rm s}(p) \le 1/(1-p)$ is easily obtained: for the mixed formula
to be satisfiable, necessarily the sub-formula obtained by retaining only
the clauses of length 2 must be satisfiable as well. In fact this bound is
tight for all values of $p\in[0,2/5]$. During the execution of the
algorithm the ratio $\alpha$ and the fraction $p$ are `dynamical' parameters,
changing with the fraction $t=T/N$ of variables assigned by the
algorithm. They define the coordinates of the representative point of
the instance at `time' $t$ in the $(p,\alpha)$ plane of Figure
\ref{fig_trajectories}. The motion of the representative point defines
the search trajectory of the algorithm. Trajectories start from the
point of coordinates $p(0)=1,\alpha(0)=\alpha$ and end up on the
$\alpha=0$ axis when a solution is found. The probability of success
is positive as long as the 2-$\sat$ subformula is satisfiable, that is,
as long as $\alpha \cdot (1-p)<1$. In other words success is possible
provided the trajectory does not cross the contradiction line 
$\alpha =1/(1-p)$ (Figure \ref{fig_trajectories}). The largest initial
ratio $\alpha$ such that no crossing occurs defines $\alpha _H$.
Notice that the search trajectory is a stochastic object. However
Franco has shown that the
deviations from its average locus in the plane vanish in the
$N\to\infty$ limit (concentration phenomenon). Large deviations from
the typical behavior can be calculated e.g. to estimate the
probability of success above $\alpha _H$~\cite{Cocp}.

The precise form of $\mathcal P_\mathrm{success}$ and the
value $\alpha_H$ of the ratio where it vanishes are specific to the heuristic
considered (UC in (\ref{p_success_UC})). However the behavior of the
probability close to $\alpha_H$ is largely independent of the
heuristic (provided it preserves the uniformity of the subformulas
generated):
\begin{eqnarray}
\ln \mathcal P_\mathrm{success}\big(\alpha = \alpha_H(1-\lambda)\big) \sim 
- \lambda ^ {-1/2}.
\end{eqnarray}
This universality can loosely be interpreted by observing that for
$\alpha$ close to $\alpha_H$ the trajectory will pass very close to
the contradiction curve $\alpha \cdot (1-p) = 1$, which characterizes the
locus of the points where the probability that a variable is assigned
by the heuristics $H$ vanishes (and all the variables are assigned by
Unit Propagation). The value of $\alpha_H$ depend on the ``shape'' of
the trajectory far from this curve, and will therefore depend on the
heuristics, but the probability of success (i.e. of avoiding the
contradiction curve) for values of $\alpha$ close to $\alpha_H$ will
only depend on the local behavior of the trajectory close to the
contradiction curve, a region where most variables are assigned
through Unit Propagation and not sensitive to the heuristics.

The finite-size corrections to equation (\ref{p_success_UC}) are also
universal (i.e. independent on the heuristics):
\begin{eqnarray} \label{sca4}
        \ln \mathcal P _\mathrm{success}(\alpha=\alpha_H(1-\lambda),N) \sim -
        N^{1/6}\;{\cal F}(\lambda N^{1/3}) \ ,
\end{eqnarray}
where ${\cal F}$ is a universal scaling function which can be 
exactly expressed in terms of the Airy function~\cite{De04}. This
result indicates that right at $\alpha_H$ the probability of success
decreases as a stretched exponential $\sim \exp (- cst\; N^{\frac 16})$.

The
exponent $\frac 13$ suggests that the critical scaling of ${\cal P}$
is related to random graphs. After $T=t\,N$ steps of the procedure, 
the sub-formula will consists of
$C_3, C_2$ and $C_1$ clauses of length 3, 2 and 1 respectively (notice
that these are \emph{extensive}, i.e. $O(N)$ quantities). We can
represent the clauses of length 1 and 2 (which are the relevant ones
to understand the generation of contradictions) as an oriented graph
$\mathcal G$ in the following way. We will have a vertex for each
literal, and represent 1-clauses by ``marking'' the literal
appearing in each; a 2-clause will be represented by two directed
edges, corresponding to the two implications equivalent to the clause
(for example, $x_1 \vee \bar x_2$ is represented by the directed edges
$\bar x_1 \rightarrow \bar x_2$ and $x_2 \rightarrow x_1$). The
average out-degree of the vertices in the graph is
$\gamma= C_2/(N-T)=\alpha(t)(1-p(t))$.

What is the effect of the algorithm on $\mathcal G$? The algorithm
will proceed in ``rounds'': a variable is set by the heuristics, and a
series of Unit Propagations are performed until no more unit clauses
are left, at which point a new round starts. Notice that during a
round, extensive quantities as $C_1,C_2,C_3$ are likely to vary by
bounded amounts and $\gamma$ to vary by $O(\frac 1N)$ 
(this is the very reason that
guarantees that these quantities are concentrated around their
mean). At each step of Unit Propagation, a marked literal (say $x$) is
assigned and removed from $\mathcal G$, together with all the edges
connected to it, and the ``descendants'' of $x$ (i.e. the literals at
the end of outgoing edges) are marked.  Also $\bar x$ is removed
together with its edges, but its descendants are not
marked. Therefore, the marked vertices ``diffuse'' in a connected
component of $\mathcal G$ following directed edges. Moreover, at each
step new edges corresponding to clauses of length 3 that get
simplified into clauses of length 2 are added to the graph.

When $\gamma > 1$, $\mathcal G$ undergoes a directed percolation
transition, and a giant component of size $O(N)$ appears, in which it
is possible to go from any vertex to any other vertex by following a
directed path. When this happens, there is a finite probability that
two opposite literals $x$ and $\bar x$ can be reached from some other
literal $y$ following a directed path. If $\bar y$ is selected by Unit
Propagation, at some time both $x$ and $\bar x$ will be marked, and
this corresponds to a contradiction. This simple argument explains
more than just the condition $\gamma = \alpha \cdot (1-p) = 1$ for the
failure of the heuristic search. It can also be used to explain the
the exponent $\frac 16$ in the scaling (\ref{sca4}), 
see \cite{De04,Mo07} for more details.

\subsection{Backtrack-based search: the Davis-Putnam-Loveland-Logeman procedure}
\label{sec_DPLL}
\index{Davis Logemann Loveland procedure}
\index{backtracking}

The heuristic search procedure of the previous Section can be easily
turned into a complete procedure for finding solutions or proving that
formulas are not satisfiable. When a contradiction is found the
algorithm now backtracks to the last assigned variable (by the
heuristic; unit clause propagations are merely consequences of previous
assignments), invert it, and the search resumes. If another
contradiction is found the algorithm backtracks to the last-but-one
assigned variable and so on. The algorithm stops either if a solution
is found or all possible backtracks have been unsuccessful and a proof of
unsatisfiability is obtained. This algorithm was proposed by Davis, Putnam, Loveland and Logemann
and is referred to as DPLL in the following.

The history of the search process can be represented by a search tree,
where the nodes represent the variables assigned, and the descending
edges their values (Figure \ref{fig-tree}).
The leaves of the tree correspond to solutions (S), or
to contradictions (C). The analysis of the $\alpha<\alpha_H$ regime in
the previous Section
leads us to the conclusion that search trees look like
Figure \ref{fig-tree}A at small ratios\footnote{A small amount of
  backtracking may be necessary to find the solution since
  ${\mathcal P}_{\rm success}<1$ \cite{Fr96}, 
but the overall picture of a single branch is not qualitatively affected.}.

For ratios $\alpha>\alpha _H$ DPLL is very likely to find a
contradiction. Backtracking enters into play, and is responsible for
the drastic slowing down of the algorithm. The success-to-failure
transition takes place in the non-backtracking algorithm into a
polynomial-to-exponential transition in DPLL. The question is to
compute the growth exponent of the average tree size, $T\sim
e^{N\tau (\alpha)}$, as a function of the ratio $\alpha$.

\subsubsection{Exponential regime: Unsatisfiable formulas}

\begin{figure}
\begin{center}
\includegraphics[width=130pt,angle=-90]{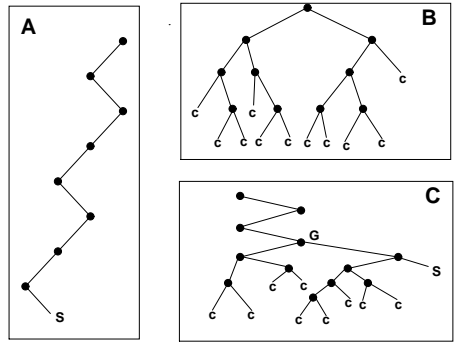}
\caption{Search trees generated by DPLL:
{\bf A.} linear, satisfiable ($\alpha < \alpha_H$); 
{\bf B.} exponential, unsatisfiable ($\alpha > \alpha _c$). 
{\bf  C.} exponential, satisfiable ($\alpha_H<\alpha<\alpha_c$);
Leaves are marked with S (solutions) or C (contradictions). G is the highest
node to which DPLL backtracks, see Figure \ref{fig_trajectories}.}
\label{fig-tree}
\end{center}
\end{figure}

Consider first the case of unsatisfiable formulas
($\alpha> \alpha_{\rm s}$) where all leaves carry contradictions after
DPLL halts (Figure \ref{fig-tree}B).
DPLL builds the tree in a sequential manner, adding nodes
and edges one after the other, and completing branches through
backtracking steps. We can think of the same search tree built in a
parallel way~\cite{Co01}. At time (depth $T$) our tree is
composed of $L(T)\le 2^T$ branches, each carrying a partial assignment
over $T$
variables. Step $T$ consists in assigning one more variable to each
branch, according to DPLL rules, that is, through unit-propagation or
the heuristic rule. In the latter case we will speak of a splitting
event, as two branches will emerge from this node, corresponding to the
two possible values of the variable assigned.
The possible consequences of this assignment are the
emergence of a contradiction (which put an end to the branch), 
or the simplification of the attached formulas (the branch keeps growing).

The number of branches $L(T)$ is a stochastic variable. Its average
value can be calculated as follows~\cite{Mon05}. Let us define the
average number $L(\vec C;T)$ of branches of depth $T$ which bear a
formula containing $C_3$ (resp. $C_2$, $C_1$) equations of length
3 (resp. 2,1), with $\vec C=(C_1,C_2,C_3)$ 
Initially $L(\vec C;0)=1$ for $\vec C=(0,0,\alpha N)$, 0 otherwise.
We shall call $M(\vec C',\vec C;T)$ the average
number of branches described by $\vec C'$ generated from a $\vec C$ branch 
once the $T^{th}$ variable is assigned~\cite{Co01,Mo07}.
We have $0\le M\le 2$, the extreme values corresponding to a
contradiction and to a split respectively. We claim that
\begin{equation}\label{evol5}
L(\vec C';T+1) = \sum _{\vec C} M(\vec C',\vec C;T) \; L(\vec C;T) \ .
\end{equation} 
Evolution equation (\ref{evol5}) could look like somewhat suspicious
at first sight due to its similarity with
the approximation we have sketched in Sec.~\ref{secrwsat2} for the analysis of
PRW$\sat$. Yet, thanks to the linearity of expectation, the
correlations between the branches (or better, the instances carried by
the branches) do not matter as far as the average number of branches
is concerned.

For large $N$ we expect that the number of alive (not hit by a contradiction)
branches grows exponentially with the depth, or, equivalently,
\begin{equation}
\sum _{C_1,C_2,C_3} L(C_1,C_2,C_3;T) \sim e ^{N \; \lambda (t) + o(N)}
\end{equation}
The argument of the exponential, $\lambda(t)$, can be found
using partial differential equation techniques generalizing the
ordinary differential equation techniques of a single branch in the
absence of backtracking (Section \ref{secuc}). Details can be found in 
\cite{Mon05}. The outcome is that $\lambda (t)$ is a function growing 
from $\lambda =0$ at
$t=0$, reaching a maximum value $\lambda_M$ for some depth $t_M$, and
decreasing at larger depths. $t_M$ is the depth in the tree
of Figure \ref{fig-tree}B where most contradictions are found; the
number of contradiction leaves is, to exponential order, $e^{N\lambda_M}$. We
conclude that the logarithm of the average size of the tree we were
looking for is
\begin{equation}
\tau  = \lambda_M \ .
\end{equation}
For large $\alpha \gg \alpha_{\rm s}$ one finds
$\tau = O(1 /\alpha)$, in 
agreement with the asymptotic scaling of~\cite{beame}. 
The calculation can be extended to higher values of $\k$.

\subsubsection{Exponential regime: Satisfiable formulas}

The above calculation holds for the unsatisfiable, exponential phase.
How can we understand the satisfiable but exponential regime
$\alpha_H<\alpha<\alpha_{\rm s}$? The resolution trajectory crosses the
$\sat$/$\unsat$ critical line $\alpha_{\rm s}(p)$
at some point G shown in Figure
\ref{fig_trajectories}. Immediately after $G$ the instance left by DPLL is
unsatisfiable. A subtree with all its leaves carrying contradictions
will develop below G (Figure \ref{fig-tree}C). The size $\tau
^G$ of this
subtree can be easily calculated from the above theory from the
knowledge of the coordinates $(p_G,\alpha_G)$ of G. 
Once this subtree has been built DPLL
backtracks to G, flips the attached variable and will finally end up
with a solution. Hence the (log of the)  number of splits necessary will be
equal to 
$\tau = (1-t_G)\, \tau_{\rm split}^G$~\cite{Co01}.
Remark that our calculation gives the logarithm of the average subtree
size starting from the typical value of G. 
Numerical experiments show that the resulting value for $\tau$ 
coincides very accurately
with the most likely tree size for finding a solution. 
The reason is that fluctuations in the
sizes are mostly due to fluctuations of the highest backtracking point
G, that is, of the first part of the search trajectory \cite{Cocp}.  

\subsection{Message passing algorithms}
\label{sec_mp}

According to the thought experiment proposed at the beginning of this Section
valuable information could be obtained from the knowledge of the marginal
probabilities of variables in the uniform measure over optimal
configurations. This is an inference problem in the graphical
model associated to the formula. In this field message passing techniques 
(for instance Belief Propagation, or the min-sum algorithm)
are widely used to compute approximately such
marginals~\cite{fgraphs,Yedidia}. These numerical procedures introduce
messages on the directed edges of the factor graph representation of the
problem (recall the definitions given in Sec.~\ref{sec_computations}), which
are iteratively updated, the new value of a message being computed from the
old values of the incoming messages (see Fig.~\ref{fig_recurs}). When the
underlying graph is a tree, the message updates are guaranteed to converge in
a finite number of steps, and provide exact results. In the presence of cycles
the convergence of these recurrence equations is not guaranteed; they can
however be used heuristically, the iterations being repeated until a fixed
point has been reached (within a tolerance threshold). Though
very few general results on the convergence in presence of loops are
known~\cite{TaJo} (see also~\cite{MoSh} for low $\alpha$ random $\sat$ formulas) 
these heuristic procedures are often found to yield good
approximation of the marginals on generic factor graph problems.

The interest in this approach for solving random $\sat$ instances was triggered
in the statistical mechanics community by the introduction of the Survey
Propagation algorithm~\cite{MeZe}. Since then several generalizations and
reinterpretations of SP have been put forward, see for
instance~\cite{BrZe,Maneva,AuKi,Pa03,SPy,reinforcement}.
In the following paragraph we present three different message passing
procedures, which differ in the nature of the messages passed between
nodes, following rather closely the presentation of~\cite{BrMeZe} to which we
refer the reader for further details. We then discuss how these procedures
have to be interleaved with assignment (decimation) steps in order to
constitute a solver algorithm. Finally we shall review results obtained in a 
particular limit case (large $\alpha$ satisfiable formulas).

\subsubsection{Definition of the message-passing algorithms}

\begin{itemize}

\item[$\bullet$] Belief Propagation (BP)

\index{message passing!belief propagation}

For the sake of readability we recall here the recursive equations
(\ref{eq_recurs}) stated in 
Sec.~\ref{sec_computations} for the uniform probability measure over the
solutions of a tree formula,
\begin{eqnarray}
h_{i \to a} &=& \sum_{b \in \partial_+ i(a)} u_{b \to i} - \sum_{b \in
\partial_- i(a)} u_{b \to i} \ , \\
u_{a \to i} &=& 
- \frac{1}{2} \ln \left(1 - \prod_{j \in \partial a \setminus i}
\frac{1-\tanh h_{j \to a}}{2} \right) \ .
\nonumber
\end{eqnarray}
where the $h$ and $u$'s messages are reals (positive for $u$), parametrizing
the marginal probabilities (beliefs) for the value of a variable in absence
of some constraint nodes around it (cf. Eq.~(\ref{eq_cavitymarginals})).
These equations can be used in the heuristic way explained above for any 
formula, and
constitute the BP message-passing equations. Note that in the course of the
simplification process the degree of the clauses change, we thus adopt here
and in the following the natural convention that sums (resp. products) over
empty sets of indices are equal to 0 (resp. 1). 

\item[$\bullet$] Warning Propagation (WP)

\index{message passing!warning propagation}

The above-stated version of the BP equations become ill-defined for an 
unsatisfiable formula, whether this 
was the case of the original formula or because of some wrong assignment
steps; in particular the normalization constant of Eq.~(\ref{eq_mu})
vanishes. A way to cure this problem consists in introducing a 
fictitious inverse temperature $\beta$ and deriving the BP equations 
corresponding to the regularized Gibbs-Boltzmann probability law (\ref{Gibbs}),
taking as the energy function the number of unsatisfied constraints. In the
limit $\beta \to \infty$, in which the Gibbs-Boltzmann measure concentrates 
on the optimal assignments, one can single out a part of the information
conveyed by the BP equations to obtain the simpler 
Warning Propagation rules. Indeed 
the messages $h,u$ are at leading order proportional to $\beta$, with
proportionality coefficients we shall denote $\widehat{h}$ and $\widehat{u}$.
These messages are less informative than the ones of BP, yet simpler to handle.
One finds indeed that instead of reals the WP messages are integers, more
precisely $\widehat{h} \in \mathbb{Z}$ and $\widehat{u} \in \{0,1\}$. They obey
the following recursive equations (with a structure similar to the ones of BP),
\begin{eqnarray}
\widehat{h}_{i \to a} &=& \sum_{b \in \partial_+ i(a)} 
\widehat{u}_{b \to i} - \sum_{b \in \partial_- i(a)} 
\widehat{u}_{b \to i} \ , \nonumber \\
\widehat{u}_{a \to i} &=& \prod_{j \in \partial a \setminus i} 
\mathbb{I}(\widehat{h}_{j \to a} < 0 ) \ ,
\label{eq_WP}
\end{eqnarray}
where $\mathbb{I}(E)$ is the indicator function of the event $E$.
The interpretation of these equations goes as follows.
$\widehat{u}_{a \to i}$ is equal to 1 if in all optimal assignments of the
amputated formula in which $i$ is only constrained by $a$, $i$ takes the
value satisfying $a$. This happens if all other variables of clause $a$ 
(i.e. $\partial a \setminus i$) are required to take their values unsatisfying
$a$, hence the form of the right part of (\ref{eq_WP}). In such a case
we say that $a$ sends a warning to variable $i$. In the first part of 
(\ref{eq_WP}), the message $\widehat{h}_{i \to a}$ sent by a variable to a 
clause is computed by pondering the number of warnings sent by all other 
clauses; it will in particular be negative if a majority of clauses requires 
$i$ to take the value unsatisfying $a$.

\item[$\bullet$] Survey Propagation (SP)

\index{message passing!survey propagation}

The convergence of BP and WP iterations is not ensured on loopy graphs. In
particular the clustering phenomenon described in Sec.~\ref{sec_clustering}
is likely to spoil the efficiency of these procedures. The Survey Propagation
(SP) algorithm introduced in~\cite{MeZe} has been designed to deal with
these clustered space of configurations. The underlying idea is that the
simple iterations (of BP or WP type) remain valid inside each cluster of
optimal assignments; for each of these clusters $\gamma$ and each directed 
edge of the factor graph one has a message $h_{i \to a}^\gamma$ (and
$u_{a \to i}^\gamma$). One introduces on each edge a survey of these messages,
defined as their probability distribution with respect to the choice of the
clusters. Then some hypotheses are made on the structure of the cluster 
decomposition in order to write closed equations on the survey. 
We explicit now this approach in a version adapted to satisfiable 
instances~\cite{BrMeZe}, taking as the basic building block the WP equations.
This leads to a rather simple form of the survey. Indeed 
$\widehat{u}_{a \to i}$ can only take two values, its probability distribution
can thus be parametrized by a single real $\delta_{a \to i} \in [0,1]$,
the probability that $\widehat{u}_{a \to i}=1$. Similarly the survey
$\gamma_{i \to a}$ is the probability that $\widehat{h}_{i \to a} < 0$.
The second part of (\ref{eq_WP}) is readily translated in probabilistic terms,
\begin{equation}
\delta_{a \to i} = \prod_{j \in \partial a
\setminus i} \gamma_{j\to a} \ .
\end{equation}
The other part of the recursion takes a slightly more complicated form,
\begin{eqnarray}
\gamma_{i \to a} &=& \frac{(1-\pi^-_{i \to a})\pi^+_{i \to a}}
{\pi^+_{i \to a} + \pi^-_{i \to a} - \pi^+_{i \to a} \pi^-_{i \to a} } \ , 
\nonumber \\
&& {\rm with} \quad
\begin{cases}
\pi^+_{i \to a} = \underset{b\in \partial_+i(a)}{\prod} (1-\delta_{b \to i}) \\
\pi^-_{i \to a} = \underset{b\in \partial_-i(a)}{\prod} (1-\delta_{b \to i})
\end{cases} \ .
\end{eqnarray}
In this equation $\pi^+_{i \to a}$ (resp. $\pi^-_{i \to a}$) corresponds to
the probability that none of the clauses agreeing (resp. disagreeing) with
$a$ on the value of the literal of $i$ sends a warning. For $i$ to be 
constrained to the value unsatisfying $a$, at least one of the clauses of 
$\partial_-i(a)$ should send a warning, and none of $\partial_+i(a)$,
which explains the form of the numerator of $\gamma_{i \to a}$. The
denominator arises from the exclusion of the event that both clauses in
$\partial_+i(a)$ and $\partial_-i(a)$ send messages, a contradictory event in
this version of SP which is devised for satisfiable formulas.

From the statistical mechanics point of view the SP equations arise from
a 1RSB cavity calculation, as sketched in Sec.~\ref{sec_computations},
in the zero temperature limit ($\beta \to \infty$) and vanishing Parisi 
parameter $m$, these two limits being either taken simultaneously as 
in~\cite{MeZe,SPy} or successively~\cite{KrMoRiSeZd}. One can thus
compute, from the solution of the recursive equations on a single formula, 
an estimation of its complexity, i.e. the number of its clusters 
(irrespectively of their sizes). The message passing procedure can also be 
adapted, at the price of technical complications, to unsatisfiable clustered
formulas~\cite{SPy}. Note also that the above SP equations have
been shown to correspond to the BP ones in an extended configuration
space where variables can take a ``joker'' value~\cite{BrZe,Maneva}, mimicking
the variables which are not frozen to a single value in all the assignments
of a given cluster. Heuristic interpolations between the BP and SP equations
have been studied in~\cite{AuKi,Maneva}.

\end{itemize}

\subsubsection{Exploiting the information}

The information provided by these message passing procedures can be exploited
in order to solve satisfiability formulas; in the algorithm sketched
at the beginning of Sec.~\ref{secuc} the heuristic choice
of the assigned variable, and its truth value, can be done according to the
results of the message passing on the current formula. If BP were an
exact inference algorithm, one could choose any unassigned variable, compute
its marginal according to Eq.~(\ref{eq_mui}), and draw it according
to this probability. Of course BP is only an approximate procedure, hence
a practical implementation of this idea should privilege the variables
with marginal probabilities closest to a deterministic law (i.e. with the 
largest $|h_i|$), motivated by the intuition that these are the least subject
to the approximation errors of BP. Similarly, if the message passing procedure
used at each assignment step is WP, one can fix the variable with the largest
$|\widehat{h}_i|$ to the value corresponding to the sign of $\widehat{h}_i$.
In the case of SP, the solution of the message passing equations are used
to compute, for each unassigned variable $i$, a triplet of numbers
$(\gamma_i^+,\gamma_i^-,\gamma_i^0)$ according to
\begin{eqnarray}
\gamma_i^+ &=& \frac{(1-\pi^+_i)\pi^-_i}
{\pi^+_i + \pi^-_i - \pi^+_i \pi^-_i } \ , \quad 
\gamma_i^- = \frac{(1-\pi^-_i)\pi^+_i}
{\pi^+_i + \pi^-_i - \pi^+_i \pi^-_i } \ , \quad 
\gamma_i^0 = 1 - \gamma_i^+ - \gamma_i^- \ , 
\nonumber\\ &&{\rm with} \quad
\begin{cases}
\pi^+_i = \underset{a\in \partial_+ i}{\prod} (1-\delta_{a \to i}) \\
\pi^-_i = \underset{a\in \partial_- i}{\prod} (1-\delta_{a \to i})
\end{cases} \ .
\end{eqnarray}
$\gamma_i^+$ (resp. $\gamma_i^-$) is interpreted as the fraction of clusters
in which $\s_i=+1$ (resp. $\s_i=-1$) in all solutions of the cluster, hence
$\gamma_i^0$ corresponds to the clusters in which $\s_i$ can take both values.
In the version of~\cite{BrMeZe}, one then choose the variable with the largest
$|\gamma_i^+ - \gamma_i^-|$, and fix it to $\s_i=+1$ (resp. $\s_i=-1$) if
$\gamma_i^+ > \gamma_i^-$ (resp. $\gamma_i^+ < \gamma_i^-$). In this way one 
tries to select an assignment preserving the maximal number of clusters.

Of course many variants of these heuristic rules can be devised; for instance
after each message passing computation one can fix a finite fraction of
the variables (instead of a single one), allows for some amount of 
backtracking~\cite{Pa03b}, or increase a soft bias instead of assigning
completely a variable~\cite{reinforcement}. Moreover the tolerance on the
level of convergence of the message passing itself can also be adjusted.
All these implementation choices will affect the performances of the solver,
in particular the maximal value of $\alpha$ up to which random $\sat$ instances
are solved efficiently, and thus makes difficult a precise statement about
the limits of these algorithms. In consequence we shall only report the
impressive result of~\cite{BrMeZe},  which presents an 
implementation~\cite{SP_implementation} working for random 3-$\sat$ instances
up to $\alpha=4.24$ (very close to the conjectured satisfiability threshold
$\alpha_{\rm s}\approx 4.267$) for problem sizes as large as $N=10^7$.

The theoretical understanding of these message passing inspired solvers is
still poor compared to the algorithms studied in Sec.~\ref{secuc},
which use much simpler heuristics in their assignment steps. One difficulty
is the description of the residual formula after an extensive number of 
variables have been assigned; because of the correlations between successive
steps of the algorithm this residual formula is not uniformly distributed
conditioned on a few dynamical parameters, as was the case with 
$(\alpha(t),p(t))$ for the simpler heuristics of Sec.~\ref{secuc}.
One version of BP guided decimation could however be studied analytically
in~\cite{Allerton}, by means of an analysis of the thought experiment
discussed at the beginning of Sec.~\ref{sec_decimation}. 
The study of another simple message passing algorithm is presented in the 
next paragraph.

\subsubsection{Warning Propagation on dense random formulas}

Feige proved in~\cite{Feige1} a remarkable connection between
the \emph{worst-case} complexity of approximation problems and the structure
of \emph{random} 3-$\sat$ at large (but independent of $N$) values
of the ratio $\alpha$. 
He introduced the following hardness hypothesis
for random 3-$\sat$ formulas:

\vskip5pt

{\rm Hypothesis 1:} {\it Even if $\alpha$ is arbitrarily large (but
  independent of $N$), there is no polynomial time algorithm that on
  most 3-SAT formulas outputs UNSAT, and always outputs SAT on a 3-SAT
  formula that is satisfiable}.

\vskip5pt 
\noindent 
and used it to derive  hardness of approximation results
for various computational problems.
As we have seen these instances are typically unsatisfiable;
the problem of interest is thus to recognize efficiently the rare satisfiable
instances of the distribution.

A variant of this problem was studied in~\cite{Feige2}, where WP
was proven to be effective in finding solutions of dense planted random
formulas (the planted distribution is the uniform distribution conditioned
on being satisfied by a given assignment). More precisely, \cite{Feige2}
proves that for $\alpha$ large enough (but
independent of $N$), the following holds with probability $1 -
e^{-O(\alpha)}$:
\begin{enumerate}
\item WP converges after at most $O(\ln N)$ iterations.
\item If a variable $i$ has $\widehat{h}_i \neq 0$, then
the sign of $\widehat{h}_i$ is equal to the value of $\s_i$ in the
planted assignment. The number of such variables is bigger than $N
(1-e^{-O(\alpha)})$ (i.e. almost all variables can be reconstructed from
the values of $\widehat{h}_i$).
\item Once these variables are fixed to their correct assignments,
the remaining formula can be satisfied in time $O(N)$ (in fact, it is
a tree formula).
\end{enumerate}
On the basis of non-rigorous statistical mechanics methods, these
results were argued in~\cite{AMZ06} to remain true when the planted 
distribution is replaced by the uniform distribution conditioned on being
satisfiable. In other words by iterating WP for a number of
iterations bigger than $O(\ln N)$ one is able to detect the rare satisfiable
instances at large $\alpha$. The argument is based on the similarity
of structure between the two distributions at large $\alpha$, namely the
existence of a single, small cluster of solutions where almost all variables
are frozen to a given value. This correspondence between the two distributions
of instances was proven rigorously in~\cite{why}, where it was also shown that
a related polynomial algorithm succeeds with high probability in finding
solutions of the satisfiable distribution of large enough density $\alpha$.

These results indicate that a stronger form of
hypothesis 1, obtained by replacing {\em always} with {\em
 with probability $p$} (with respect to the uniform distribution over the
formulas and possibly to some randomness built in the algorithm), is wrong
for any $p<1$.
However, the validity of hypothesis 1 is still unknown for random
3-$\sat$ instances.
Nevertheless, this result is interesting because it is one of the rare cases
in which the performances of a message-passing algorithm could be analyzed in
full detail.

\section{Conclusion}
\label{sec_conclu}

This review was mainly dedicated to the random $\k$-Satisfiability and
$\k$-Xor-Satisfiability problems; the approach and results we presented
however extend to other random decision problems, in particular
random graph $q$-coloring. This problem consists in deciding whether
each vertex of a graph can be assigned one out of $q$ possible colors,
without giving the same color to the two extremities of an edge. 
When input graphs are randomly drawn from Erd\"os-Renyi (ER) ensemble 
$G(N,p=c/N)$ a phase diagram similar to the one of $k$-SAT 
(Section \ref{sec:phase_transitions}) 
is obtained. There exists a colorable/uncolorable phase transition for
some critical average degree $c_{\rm s}(q)$, with for instance
$c_{\rm s}(3)\simeq 4.69$~\cite{col1}. The colorable phase also exhibits 
the clustering and condensation transitions~\cite{col2} we explained
on the example of the $\k$-Satisfiability.
Actually what seems to matter here is rather the structure of inputs and
the symmetry properties of the decision problem rather than its specific 
details. All the above considered input 
models share a common, underlying ER random graph structure. 
From this point of view it would be interesting to `escape' from the 
ER ensemble and consider more structured graphs e.g. embedded in a low
dimensional space.

To what extent the similarity between phase diagrams correspond to
similar behaviour in terms of hardness of resolution is an open
question. Consider the case of rare satisfiable instances for the 
random $k$-SAT and $k$-XORSAT well above their sat/unsat thresholds
(Section~\ref{sec_decimation}).  
Both problems share very similar statistical features. 
However, while a simple message-passing algorithm allows one to
easily find a (the) solution for the $k$-SAT problem this algorithm is
inefficient for random $k$-XORSAT. Actually the local or
decimation-based algorithms of Sections~\ref{sec_localsearch} and
\ref{sec_decimation} are efficient to find
solution to rare satisfable instances of random $\k$-SAT 
\cite{Ba02}, but none of
them works for random $\k$-XORSAT (while the problem is in P!).  
This example raises the important question of
the relationship between the statistical properties of solutions (or
quasi-solutions) encoded in the phase diagram and the (average) 
computational hardness. Very little is known about this crucial point;
on intuitive grounds one could expect the clustering phenomenon to prevent
an efficient solving of formulas by local search algorithms of the 
random walk type. This is indeed true for a particular class of stochastic
processes~\cite{MoSe2}, those which respect the so-called detailed balance 
conditions. This connection between clustering and hardness of resolution for
local search algorithms is much less obvious when the detailed balance 
conditions are not respected, which is the case for most of the efficient
variants of PRWSAT. 


\begin{thebibliography}{101}
\expandafter\ifx\csname natexlab\endcsname\relax\def\natexlab#1{#1}\fi
\expandafter\ifx\csname bibnamefont\endcsname\relax
  \def\bibnamefont#1{#1}\fi
\expandafter\ifx\csname bibfnamefont\endcsname\relax
  \def\bibfnamefont#1{#1}\fi
\expandafter\ifx\csname citenamefont\endcsname\relax
  \def\citenamefont#1{#1}\fi
\expandafter\ifx\csname url\endcsname\relax
  \def\url#1{\texttt{#1}}\fi
\expandafter\ifx\csname urlprefix\endcsname\relax\def\urlprefix{URL }\fi
\providecommand{\bibinfo}[2]{#2}
\providecommand{\eprint}[2][]{\url{#2}}

\bibitem[{\citenamefont{M\'ezard et~al.}(1987)\citenamefont{M\'ezard, Parisi,
  and Virasoro}}]{Beyond}
\bibinfo{author}{\bibfnamefont{M.}~\bibnamefont{M\'ezard}},
  \bibinfo{author}{\bibfnamefont{G.}~\bibnamefont{Parisi}}, \bibnamefont{and}
  \bibinfo{author}{\bibfnamefont{M.}~\bibnamefont{Virasoro}},
  \emph{\bibinfo{title}{Spin glass theory and beyond}}
  (\bibinfo{publisher}{World Scientific}, \bibinfo{address}{Singapore},
  \bibinfo{year}{1987}).

\bibitem[{\citenamefont{Papadimitriou and Steiglitz}(1998)}]{Pa83}
\bibinfo{author}{\bibfnamefont{C.}~\bibnamefont{Papadimitriou}}
  \bibnamefont{and}
  \bibinfo{author}{\bibfnamefont{K.}~\bibnamefont{Steiglitz}},
  \emph{\bibinfo{title}{Combinatorial Optimization: Algorithms and Complexity}}
  (\bibinfo{publisher}{Dover}, \bibinfo{address}{New York},
  \bibinfo{year}{1998}).

\bibitem[{\citenamefont{Fu and Anderson}(1986)}]{Fu85}
\bibinfo{author}{\bibfnamefont{Y.}~\bibnamefont{Fu}} \bibnamefont{and}
  \bibinfo{author}{\bibfnamefont{P.~W.} \bibnamefont{Anderson}},
  \bibinfo{journal}{Journal of Physics A: Mathematical and General}
  \textbf{\bibinfo{volume}{19}}, \bibinfo{pages}{1605} (\bibinfo{year}{1986}).

\bibitem[{\citenamefont{Mitchell et~al.}(1992)\citenamefont{Mitchell, Selman,
  and Levesque}}]{transition_exp}
\bibinfo{author}{\bibfnamefont{D.}~\bibnamefont{Mitchell}},
  \bibinfo{author}{\bibfnamefont{B.}~\bibnamefont{Selman}}, \bibnamefont{and}
  \bibinfo{author}{\bibfnamefont{H.}~\bibnamefont{Levesque}}
  (\bibinfo{year}{1992}), no. \bibinfo{number}{459} in
  \bibinfo{series}{Proceedings of the Tenth National Conference on Artificial
  Intelligence}.

\bibitem[{\citenamefont{Hertz et~al.}(1991)\citenamefont{Hertz, Krogh, and
  Palmer}}]{revue-perceptron}
\bibinfo{author}{\bibfnamefont{J.}~\bibnamefont{Hertz}},
  \bibinfo{author}{\bibfnamefont{A.}~\bibnamefont{Krogh}}, \bibnamefont{and}
  \bibinfo{author}{\bibfnamefont{R.}~\bibnamefont{Palmer}},
  \emph{\bibinfo{title}{Introduction to the theory of neural computation}},
  Santa Fe Institute Studies in the Science of Complexity
  (\bibinfo{publisher}{Addison-Wesley}, \bibinfo{address}{Redwood city (CA)},
  \bibinfo{year}{1991}).

\bibitem[{\citenamefont{Cover}(1965)}]{Cover}
\bibinfo{author}{\bibfnamefont{T.}~\bibnamefont{Cover}}, \bibinfo{journal}{IEEE
  Transactions on Electronic Computers} \textbf{\bibinfo{volume}{14}},
  \bibinfo{pages}{326} (\bibinfo{year}{1965}).

\bibitem[{\citenamefont{Janson et~al.}(2000)\citenamefont{Janson, Luczak, and
  Rucinski}}]{random_graphs}
\bibinfo{author}{\bibfnamefont{S.}~\bibnamefont{Janson}},
  \bibinfo{author}{\bibfnamefont{T.}~\bibnamefont{Luczak}}, \bibnamefont{and}
  \bibinfo{author}{\bibfnamefont{A.}~\bibnamefont{Rucinski}},
  \emph{\bibinfo{title}{Random graphs}} (\bibinfo{publisher}{John Wiley and
  Sons}, \bibinfo{address}{New York}, \bibinfo{year}{2000}).

\bibitem[{\citenamefont{Friedgut}(1999)}]{Friedgut}
\bibinfo{author}{\bibfnamefont{E.}~\bibnamefont{Friedgut}},
  \bibinfo{journal}{Journal of the American Mathematical Society}
  \textbf{\bibinfo{volume}{12}}, \bibinfo{pages}{1017} (\bibinfo{year}{1999}).

\bibitem[{\citenamefont{Dubois}(2001)}]{transition_ub}
\bibinfo{author}{\bibfnamefont{O.}~\bibnamefont{Dubois}},
  \bibinfo{journal}{Theoret. Comput. Sci.} \textbf{\bibinfo{volume}{265}},
  \bibinfo{pages}{187} (\bibinfo{year}{2001}).

\bibitem[{\citenamefont{Franco}(2001)}]{transition_lb}
\bibinfo{author}{\bibfnamefont{J.}~\bibnamefont{Franco}},
  \bibinfo{journal}{Theoret. Comput. Sci.} \textbf{\bibinfo{volume}{265}},
  \bibinfo{pages}{147} (\bibinfo{year}{2001}).

\bibitem[{\citenamefont{Achlioptas and Peres}(2004)}]{transition_largek}
\bibinfo{author}{\bibfnamefont{D.}~\bibnamefont{Achlioptas}} \bibnamefont{and}
  \bibinfo{author}{\bibfnamefont{Y.}~\bibnamefont{Peres}},
  \bibinfo{journal}{Journal of the American Mathematical Society}
  \textbf{\bibinfo{volume}{17}}, \bibinfo{pages}{947} (\bibinfo{year}{2004}).

\bibitem[{cha()}]{chapter_randomsat}
\emph{\bibinfo{title}{Chapter random sat, this volume}}.

\bibitem[{\citenamefont{Alon and Spencer}(2000)}]{2nd_moment}
\bibinfo{author}{\bibfnamefont{N.}~\bibnamefont{Alon}} \bibnamefont{and}
  \bibinfo{author}{\bibfnamefont{J.}~\bibnamefont{Spencer}},
  \emph{\bibinfo{title}{The probabilistic method}} (\bibinfo{publisher}{John
  Wiley and sons}, \bibinfo{address}{New York}, \bibinfo{year}{2000}).

\bibitem[{\citenamefont{Dembo and Zeitouni}(1998)}]{ldp}
\bibinfo{author}{\bibfnamefont{A.}~\bibnamefont{Dembo}} \bibnamefont{and}
  \bibinfo{author}{\bibfnamefont{O.}~\bibnamefont{Zeitouni}},
  \emph{\bibinfo{title}{Large deviations. Theory and applications}}
  (\bibinfo{publisher}{Springer}, \bibinfo{address}{Berlin},
  \bibinfo{year}{1998}).

\bibitem[{\citenamefont{Krauth and Mezard}(1989)}]{kra89}
\bibinfo{author}{\bibfnamefont{W.}~\bibnamefont{Krauth}} \bibnamefont{and}
  \bibinfo{author}{\bibfnamefont{M.}~\bibnamefont{Mezard}},
  \bibinfo{journal}{J. Physique} \textbf{\bibinfo{volume}{50}},
  \bibinfo{pages}{3057} (\bibinfo{year}{1989}).

\bibitem[{\citenamefont{Ma}(1985)}]{statmech1}
\bibinfo{author}{\bibfnamefont{S.~K.} \bibnamefont{Ma}},
  \emph{\bibinfo{title}{Statistical Mechanics}} (\bibinfo{publisher}{World
  Scientific}, \bibinfo{address}{Singapore}, \bibinfo{year}{1985}).

\bibitem[{\citenamefont{Huang}(1990)}]{statmech2}
\bibinfo{author}{\bibfnamefont{K.}~\bibnamefont{Huang}},
  \emph{\bibinfo{title}{Statistical Mechanics}} (\bibinfo{publisher}{John Wiley
  and Sons}, \bibinfo{address}{New York}, \bibinfo{year}{1990}).

\bibitem[{\citenamefont{Broder et~al.}(1993)\citenamefont{Broder, Frieze, and
  Upfal}}]{purelit}
\bibinfo{author}{\bibfnamefont{A.}~\bibnamefont{Broder}},
  \bibinfo{author}{\bibfnamefont{A.}~\bibnamefont{Frieze}}, \bibnamefont{and}
  \bibinfo{author}{\bibfnamefont{E.}~\bibnamefont{Upfal}}
  (\bibinfo{year}{1993}), no. \bibinfo{number}{322} in
  \bibinfo{series}{Proceedings of the Fourth Annual ACM-SIAM Symposium on
  Discrete Algorithms}.

\bibitem[{\citenamefont{Monasson and Zecchina}(1997)}]{MoZe}
\bibinfo{author}{\bibfnamefont{R.}~\bibnamefont{Monasson}} \bibnamefont{and}
  \bibinfo{author}{\bibfnamefont{R.}~\bibnamefont{Zecchina}},
  \bibinfo{journal}{Phys. Rev. E} \textbf{\bibinfo{volume}{56}},
  \bibinfo{pages}{1357} (\bibinfo{year}{1997}).

\bibitem[{\citenamefont{Biroli et~al.}(2000)\citenamefont{Biroli, Monasson, and
  Weigt}}]{BiMoWe}
\bibinfo{author}{\bibfnamefont{G.}~\bibnamefont{Biroli}},
  \bibinfo{author}{\bibfnamefont{R.}~\bibnamefont{Monasson}}, \bibnamefont{and}
  \bibinfo{author}{\bibfnamefont{M.}~\bibnamefont{Weigt}},
  \bibinfo{journal}{Eur. Phys. J. B} \textbf{\bibinfo{volume}{14}},
  \bibinfo{pages}{551} (\bibinfo{year}{2000}).

\bibitem[{\citenamefont{M\'ezard and Zecchina}(2002)}]{MeZe}
\bibinfo{author}{\bibfnamefont{M.}~\bibnamefont{M\'ezard}} \bibnamefont{and}
  \bibinfo{author}{\bibfnamefont{R.}~\bibnamefont{Zecchina}},
  \bibinfo{journal}{Phys. Rev. E} \textbf{\bibinfo{volume}{66}},
  \bibinfo{pages}{056126} (\bibinfo{year}{2002}).

\bibitem[{\citenamefont{Krzakala et~al.}(2007)\citenamefont{Krzakala,
  Montanari, Ricci-Tersenghi, Semerjian, and Zdeborova}}]{KrMoRiSeZd}
\bibinfo{author}{\bibfnamefont{F.}~\bibnamefont{Krzakala}},
  \bibinfo{author}{\bibfnamefont{A.}~\bibnamefont{Montanari}},
  \bibinfo{author}{\bibfnamefont{F.}~\bibnamefont{Ricci-Tersenghi}},
  \bibinfo{author}{\bibfnamefont{G.}~\bibnamefont{Semerjian}},
  \bibnamefont{and}
  \bibinfo{author}{\bibfnamefont{L.}~\bibnamefont{Zdeborova}},
  \bibinfo{journal}{Proceedings of the National Academy of Sciences}
  \textbf{\bibinfo{volume}{104}}, \bibinfo{pages}{10318}
  (\bibinfo{year}{2007}),
  \eprint{http://www.pnas.org/cgi/reprint/104/25/10318.pdf}.

\bibitem[{\citenamefont{Monasson and O'Kane}(1994)}]{neural_networks}
\bibinfo{author}{\bibfnamefont{R.}~\bibnamefont{Monasson}} \bibnamefont{and}
  \bibinfo{author}{\bibfnamefont{D.}~\bibnamefont{O'Kane}},
  \bibinfo{journal}{Europhysics Letters} \textbf{\bibinfo{volume}{27}},
  \bibinfo{pages}{85} (\bibinfo{year}{1994}).

\bibitem[{\citenamefont{Kirkpatrick and Thirumalai}(1987)}]{pspin}
\bibinfo{author}{\bibfnamefont{T.~R.} \bibnamefont{Kirkpatrick}}
  \bibnamefont{and}
  \bibinfo{author}{\bibfnamefont{D.}~\bibnamefont{Thirumalai}},
  \bibinfo{journal}{Phys. Rev. B} \textbf{\bibinfo{volume}{36}},
  \bibinfo{pages}{5388} (\bibinfo{year}{1987}).

\bibitem[{\citenamefont{Talagrand}(2003)}]{Talagrand_book}
\bibinfo{author}{\bibfnamefont{M.}~\bibnamefont{Talagrand}},
  \emph{\bibinfo{title}{Spin glasses: a challenge for mathematicians}}
  (\bibinfo{publisher}{Springer}, \bibinfo{address}{Berlin},
  \bibinfo{year}{2003}).

\bibitem[{\citenamefont{Panchenko and Talagrand}(2004)}]{PaTa}
\bibinfo{author}{\bibfnamefont{D.}~\bibnamefont{Panchenko}} \bibnamefont{and}
  \bibinfo{author}{\bibfnamefont{M.}~\bibnamefont{Talagrand}},
  \bibinfo{journal}{Probab. Theory Relat. Fields}
  \textbf{\bibinfo{volume}{130}}, \bibinfo{pages}{319} (\bibinfo{year}{2004}).

\bibitem[{\citenamefont{Franz and Leone}(2003)}]{FrLe}
\bibinfo{author}{\bibfnamefont{S.}~\bibnamefont{Franz}} \bibnamefont{and}
  \bibinfo{author}{\bibfnamefont{M.}~\bibnamefont{Leone}}, \bibinfo{journal}{J.
  Stat. Phys.} \textbf{\bibinfo{volume}{111}}, \bibinfo{pages}{535}
  (\bibinfo{year}{2003}).

\bibitem[{\citenamefont{M\'ezard et~al.}(2003)\citenamefont{M\'ezard,
  Ricci-Tersenghi, and Zecchina}}]{xor_1}
\bibinfo{author}{\bibfnamefont{M.}~\bibnamefont{M\'ezard}},
  \bibinfo{author}{\bibfnamefont{F.}~\bibnamefont{Ricci-Tersenghi}},
  \bibnamefont{and} \bibinfo{author}{\bibfnamefont{R.}~\bibnamefont{Zecchina}},
  \bibinfo{journal}{J. Stat. Phys.} \textbf{\bibinfo{volume}{111}},
  \bibinfo{pages}{505} (\bibinfo{year}{2003}).

\bibitem[{\citenamefont{Cocco et~al.}(2003)\citenamefont{Cocco, Dubois,
  Mandler, and Monasson}}]{xor_2}
\bibinfo{author}{\bibfnamefont{S.}~\bibnamefont{Cocco}},
  \bibinfo{author}{\bibfnamefont{O.}~\bibnamefont{Dubois}},
  \bibinfo{author}{\bibfnamefont{J.}~\bibnamefont{Mandler}}, \bibnamefont{and}
  \bibinfo{author}{\bibfnamefont{R.}~\bibnamefont{Monasson}},
  \bibinfo{journal}{Phys. Rev. Lett.} \textbf{\bibinfo{volume}{90}},
  \bibinfo{pages}{047205} (\bibinfo{year}{2003}).

\bibitem[{\citenamefont{M\'{e}zard
  et~al.}(2005{\natexlab{a}})\citenamefont{M\'{e}zard, Mora, and
  Zecchina}}]{clus_rig_xsat1}
\bibinfo{author}{\bibfnamefont{M.}~\bibnamefont{M\'{e}zard}},
  \bibinfo{author}{\bibfnamefont{T.}~\bibnamefont{Mora}}, \bibnamefont{and}
  \bibinfo{author}{\bibfnamefont{R.}~\bibnamefont{Zecchina}},
  \bibinfo{journal}{Physical Review Letters} \textbf{\bibinfo{volume}{94}},
  \bibinfo{eid}{197205} (pages~\bibinfo{numpages}{4})
  (\bibinfo{year}{2005}{\natexlab{a}}).

\bibitem[{\citenamefont{Daud\'e et~al.}(2005)\citenamefont{Daud\'e, M\'ezard,
  Mora, and Zecchina}}]{clus_rig_xsat2}
\bibinfo{author}{\bibfnamefont{H.}~\bibnamefont{Daud\'e}},
  \bibinfo{author}{\bibfnamefont{M.}~\bibnamefont{M\'ezard}},
  \bibinfo{author}{\bibfnamefont{T.}~\bibnamefont{Mora}}, \bibnamefont{and}
  \bibinfo{author}{\bibfnamefont{R.}~\bibnamefont{Zecchina}}
  (\bibinfo{year}{2005}), \bibinfo{note}{{\tt arXiv:cond-mat/0506053}}.

\bibitem[{\citenamefont{Achlioptas and Ricci-Tersenghi}(2006)}]{clus_rig_Fede}
\bibinfo{author}{\bibfnamefont{D.}~\bibnamefont{Achlioptas}} \bibnamefont{and}
  \bibinfo{author}{\bibfnamefont{F.}~\bibnamefont{Ricci-Tersenghi}},
  \bibinfo{journal}{Proceedings of the thirty-eighth annual ACM symposium on
  Theory of computing}  (\bibinfo{year}{2006}), \bibinfo{note}{{\tt
  arXiv:cs.CC/0611052}}.

\bibitem[{\citenamefont{Ricci-Tersenghi
  et~al.}(2001)\citenamefont{Ricci-Tersenghi, Weigt, and
  Zecchina}}]{xor_replica}
\bibinfo{author}{\bibfnamefont{F.}~\bibnamefont{Ricci-Tersenghi}},
  \bibinfo{author}{\bibfnamefont{M.}~\bibnamefont{Weigt}}, \bibnamefont{and}
  \bibinfo{author}{\bibfnamefont{R.}~\bibnamefont{Zecchina}},
  \bibinfo{journal}{Phys. Rev. E} \textbf{\bibinfo{volume}{63}},
  \bibinfo{pages}{026702} (\bibinfo{year}{2001}).

\bibitem[{\citenamefont{Pittel et~al.}(1996)\citenamefont{Pittel, Spencer, and
  Wormald}}]{q_core_graphs}
\bibinfo{author}{\bibfnamefont{B.}~\bibnamefont{Pittel}},
  \bibinfo{author}{\bibfnamefont{J.}~\bibnamefont{Spencer}}, \bibnamefont{and}
  \bibinfo{author}{\bibfnamefont{N.}~\bibnamefont{Wormald}},
  \bibinfo{journal}{J. Comb. Theory, Ser. B} \textbf{\bibinfo{volume}{67}},
  \bibinfo{pages}{111} (\bibinfo{year}{1996}).

\bibitem[{\citenamefont{Kurtz}(1970)}]{diff_eq}
\bibinfo{author}{\bibfnamefont{T.}~\bibnamefont{Kurtz}}, \bibinfo{journal}{J.
  Appl. Probab.} \textbf{\bibinfo{volume}{7}}, \bibinfo{pages}{49}
  (\bibinfo{year}{1970}).

\bibitem[{\citenamefont{Montanari and Semerjian}(2006{\natexlab{a}})}]{MoSe}
\bibinfo{author}{\bibfnamefont{A.}~\bibnamefont{Montanari}} \bibnamefont{and}
  \bibinfo{author}{\bibfnamefont{G.}~\bibnamefont{Semerjian}},
  \bibinfo{journal}{J. Stat. Phys.} \textbf{\bibinfo{volume}{124}},
  \bibinfo{pages}{103} (\bibinfo{year}{2006}{\natexlab{a}}).

\bibitem[{\citenamefont{Mora and M\'{e}zard}(2006)}]{clus_xxorsat}
\bibinfo{author}{\bibfnamefont{T.}~\bibnamefont{Mora}} \bibnamefont{and}
  \bibinfo{author}{\bibfnamefont{M.}~\bibnamefont{M\'{e}zard}},
  \bibinfo{journal}{Journal of Statistical Mechanics: Theory and Experiment}
  \textbf{\bibinfo{volume}{2006}}, \bibinfo{pages}{P10007}
  (\bibinfo{year}{2006}).

\bibitem[{\citenamefont{Mertens et~al.}(2006)\citenamefont{Mertens, M\'ezard,
  and Zecchina}}]{MeMeZe}
\bibinfo{author}{\bibfnamefont{S.}~\bibnamefont{Mertens}},
  \bibinfo{author}{\bibfnamefont{M.}~\bibnamefont{M\'ezard}}, \bibnamefont{and}
  \bibinfo{author}{\bibfnamefont{R.}~\bibnamefont{Zecchina}},
  \bibinfo{journal}{Random Struct. Algorithms} \textbf{\bibinfo{volume}{28}},
  \bibinfo{pages}{340} (\bibinfo{year}{2006}).

\bibitem[{\citenamefont{M\'{e}zard
  et~al.}(2005{\natexlab{b}})\citenamefont{M\'{e}zard, Palassini, and
  Rivoire}}]{MePaRi}
\bibinfo{author}{\bibfnamefont{M.}~\bibnamefont{M\'{e}zard}},
  \bibinfo{author}{\bibfnamefont{M.}~\bibnamefont{Palassini}},
  \bibnamefont{and} \bibinfo{author}{\bibfnamefont{O.}~\bibnamefont{Rivoire}},
  \bibinfo{journal}{Physical Review Letters} \textbf{\bibinfo{volume}{95}},
  \bibinfo{eid}{200202} (pages~\bibinfo{numpages}{4})
  (\bibinfo{year}{2005}{\natexlab{b}}).

\bibitem[{\citenamefont{Montanari et~al.}(2004)\citenamefont{Montanari, Parisi,
  and Ricci-Tersenghi}}]{MoPaRi}
\bibinfo{author}{\bibfnamefont{A.}~\bibnamefont{Montanari}},
  \bibinfo{author}{\bibfnamefont{G.}~\bibnamefont{Parisi}}, \bibnamefont{and}
  \bibinfo{author}{\bibfnamefont{F.}~\bibnamefont{Ricci-Tersenghi}},
  \bibinfo{journal}{Journal of Physics A: Mathematical and General}
  \textbf{\bibinfo{volume}{37}}, \bibinfo{pages}{2073} (\bibinfo{year}{2004}).

\bibitem[{\citenamefont{Mora and Zdeborova}(2007)}]{rcm}
\bibinfo{author}{\bibfnamefont{T.}~\bibnamefont{Mora}} \bibnamefont{and}
  \bibinfo{author}{\bibfnamefont{L.}~\bibnamefont{Zdeborova}}
  (\bibinfo{year}{2007}), \bibinfo{note}{{\tt arXiv:0710.3804}}.

\bibitem[{\citenamefont{Semerjian}(2008)}]{rearr_csp}
\bibinfo{author}{\bibfnamefont{G.}~\bibnamefont{Semerjian}},
  \bibinfo{journal}{J.Stat.Phys.} \textbf{\bibinfo{volume}{130}},
  \bibinfo{pages}{251} (\bibinfo{year}{2008}).

\bibitem[{\citenamefont{Monasson}(1998)}]{replica_diluted}
\bibinfo{author}{\bibfnamefont{R.}~\bibnamefont{Monasson}},
  \bibinfo{journal}{Journal of Physics A: Mathematical and General}
  \textbf{\bibinfo{volume}{31}}, \bibinfo{pages}{513} (\bibinfo{year}{1998}).

\bibitem[{\citenamefont{M\'ezard and Parisi}(2001)}]{cavity}
\bibinfo{author}{\bibfnamefont{M.}~\bibnamefont{M\'ezard}} \bibnamefont{and}
  \bibinfo{author}{\bibfnamefont{G.}~\bibnamefont{Parisi}},
  \bibinfo{journal}{Eur. Phys. J. B} \textbf{\bibinfo{volume}{20}},
  \bibinfo{pages}{217} (\bibinfo{year}{2001}).

\bibitem[{\citenamefont{M\'ezard and Parisi}(2003)}]{cavity_T0}
\bibinfo{author}{\bibfnamefont{M.}~\bibnamefont{M\'ezard}} \bibnamefont{and}
  \bibinfo{author}{\bibfnamefont{G.}~\bibnamefont{Parisi}},
  \bibinfo{journal}{J. Stat. Phys.} \textbf{\bibinfo{volume}{111}},
  \bibinfo{pages}{1} (\bibinfo{year}{2003}).

\bibitem[{\citenamefont{Kschischang et~al.}(2001)\citenamefont{Kschischang,
  Frey, and Loeliger}}]{fgraphs}
\bibinfo{author}{\bibfnamefont{F.~R.} \bibnamefont{Kschischang}},
  \bibinfo{author}{\bibfnamefont{B.~J.} \bibnamefont{Frey}}, \bibnamefont{and}
  \bibinfo{author}{\bibfnamefont{H.-A.} \bibnamefont{Loeliger}},
  \bibinfo{journal}{IEEE Trans. Inf. Theory} \textbf{\bibinfo{volume}{47}},
  \bibinfo{pages}{498} (\bibinfo{year}{2001}).

\bibitem[{\citenamefont{Braunstein et~al.}(2005)\citenamefont{Braunstein,
  M\'ezard, and Zecchina}}]{BrMeZe}
\bibinfo{author}{\bibfnamefont{A.}~\bibnamefont{Braunstein}},
  \bibinfo{author}{\bibfnamefont{M.}~\bibnamefont{M\'ezard}}, \bibnamefont{and}
  \bibinfo{author}{\bibfnamefont{R.}~\bibnamefont{Zecchina}},
  \bibinfo{journal}{Random Struct. Algorithms} \textbf{\bibinfo{volume}{27}},
  \bibinfo{pages}{201} (\bibinfo{year}{2005}).

\bibitem[{\citenamefont{Yedidia et~al.}(2001)\citenamefont{Yedidia, Freeman,
  and Weiss}}]{Yedidia}
\bibinfo{author}{\bibfnamefont{J.~S.} \bibnamefont{Yedidia}},
  \bibinfo{author}{\bibfnamefont{W.~T.} \bibnamefont{Freeman}},
  \bibnamefont{and} \bibinfo{author}{\bibfnamefont{Y.}~\bibnamefont{Weiss}},
  \bibinfo{journal}{Advances in Neural Information Processing Systems}
  \textbf{\bibinfo{volume}{13}}, \bibinfo{pages}{689} (\bibinfo{year}{2001}).

\bibitem[{\citenamefont{Yedidia et~al.}(2003)\citenamefont{Yedidia, Freeman,
  and Weiss}}]{Yedidia2}
\bibinfo{author}{\bibfnamefont{J.~S.} \bibnamefont{Yedidia}},
  \bibinfo{author}{\bibfnamefont{W.~T.} \bibnamefont{Freeman}},
  \bibnamefont{and} \bibinfo{author}{\bibfnamefont{Y.}~\bibnamefont{Weiss}}, in
  \emph{\bibinfo{booktitle}{Exploring Artificial Intelligence in the New
  Millennium}} (\bibinfo{year}{2003}), p. \bibinfo{pages}{239}.

\bibitem[{\citenamefont{Fernandez de~la Vega}(2001)}]{transition_k2}
\bibinfo{author}{\bibfnamefont{W.}~\bibnamefont{Fernandez de~la Vega}},
  \bibinfo{journal}{Theor. Comput. Sci.} \textbf{\bibinfo{volume}{265}},
  \bibinfo{pages}{131} (\bibinfo{year}{2001}).

\bibitem[{\citenamefont{Bollob\'as et~al.}(2001)\citenamefont{Bollob\'as,
  Borgs, Chayes, Kim, and Wilson}}]{FSS_k2}
\bibinfo{author}{\bibfnamefont{B.}~\bibnamefont{Bollob\'as}},
  \bibinfo{author}{\bibfnamefont{C.}~\bibnamefont{Borgs}},
  \bibinfo{author}{\bibfnamefont{J.~T.} \bibnamefont{Chayes}},
  \bibinfo{author}{\bibfnamefont{J.~H.} \bibnamefont{Kim}}, \bibnamefont{and}
  \bibinfo{author}{\bibfnamefont{D.~B.} \bibnamefont{Wilson}},
  \bibinfo{journal}{Random Struct. Algorithms} \textbf{\bibinfo{volume}{18}},
  \bibinfo{pages}{201} (\bibinfo{year}{2001}).

\bibitem[{\citenamefont{Amraoui et~al.}(2004)\citenamefont{Amraoui, Montanari,
  Richardson, and Urbanke}}]{FSS_codes}
\bibinfo{author}{\bibfnamefont{A.}~\bibnamefont{Amraoui}},
  \bibinfo{author}{\bibfnamefont{A.}~\bibnamefont{Montanari}},
  \bibinfo{author}{\bibfnamefont{T.}~\bibnamefont{Richardson}},
  \bibnamefont{and} \bibinfo{author}{\bibfnamefont{R.}~\bibnamefont{Urbanke}},
  \bibinfo{journal}{{\tt arXiv:cs.IT/0406050}}  (\bibinfo{year}{2004}).

\bibitem[{\citenamefont{Dembo and Montanari}(2007)}]{FSS_cores}
\bibinfo{author}{\bibfnamefont{A.}~\bibnamefont{Dembo}} \bibnamefont{and}
  \bibinfo{author}{\bibfnamefont{A.}~\bibnamefont{Montanari}},
  \bibinfo{journal}{{\tt arXiv:math.PR/0702007}}  (\bibinfo{year}{2007}).

\bibitem[{\citenamefont{Wilson}(2002)}]{FSS_Wilson}
\bibinfo{author}{\bibfnamefont{D.~B.} \bibnamefont{Wilson}},
  \bibinfo{journal}{Random Struct. Algorithms} \textbf{\bibinfo{volume}{21}},
  \bibinfo{pages}{182} (\bibinfo{year}{2002}).

\bibitem[{\citenamefont{Kirkpatrick and Selman}(1994)}]{FSS_KiSe}
\bibinfo{author}{\bibfnamefont{S.}~\bibnamefont{Kirkpatrick}} \bibnamefont{and}
  \bibinfo{author}{\bibfnamefont{B.}~\bibnamefont{Selman}},
  \bibinfo{journal}{Science} \textbf{\bibinfo{volume}{264}},
  \bibinfo{pages}{1297} (\bibinfo{year}{1994}).

\bibitem[{\citenamefont{Monasson et~al.}(1999)\citenamefont{Monasson, Zecchina,
  Kirkpatrick, Selman, and Troyansky}}]{FSS_Moetal}
\bibinfo{author}{\bibfnamefont{R.}~\bibnamefont{Monasson}},
  \bibinfo{author}{\bibfnamefont{R.}~\bibnamefont{Zecchina}},
  \bibinfo{author}{\bibfnamefont{S.}~\bibnamefont{Kirkpatrick}},
  \bibinfo{author}{\bibfnamefont{B.}~\bibnamefont{Selman}}, \bibnamefont{and}
  \bibinfo{author}{\bibfnamefont{L.}~\bibnamefont{Troyansky}},
  \bibinfo{journal}{Random Struct. Algorithms} \textbf{\bibinfo{volume}{15}},
  \bibinfo{pages}{414} (\bibinfo{year}{1999}).

\bibitem[{\citenamefont{De~Gregorio et~al.}(2005)\citenamefont{De~Gregorio,
  Lawlor, Bradley, and Dawson}}]{bootstrap}
\bibinfo{author}{\bibfnamefont{P.}~\bibnamefont{De~Gregorio}},
  \bibinfo{author}{\bibfnamefont{A.}~\bibnamefont{Lawlor}},
  \bibinfo{author}{\bibfnamefont{P.}~\bibnamefont{Bradley}}, \bibnamefont{and}
  \bibinfo{author}{\bibfnamefont{K.}~\bibnamefont{Dawson}},
  \bibinfo{journal}{PNAS} \textbf{\bibinfo{volume}{102}}, \bibinfo{pages}{5669}
  (\bibinfo{year}{2005}).

\bibitem[{\citenamefont{Cugliandolo}(2003)}]{Leticia}
\bibinfo{author}{\bibfnamefont{L.}~\bibnamefont{Cugliandolo}}, in
  \emph{\bibinfo{booktitle}{Slow relaxations and nonequilibrium dynamics in
  condensed matter}}, edited by \bibinfo{editor}{\bibfnamefont{J.~L.}
  \bibnamefont{Barrat}},
  \bibinfo{editor}{\bibfnamefont{M.}~\bibnamefont{Feigelman}},
  \bibinfo{editor}{\bibfnamefont{J.}~\bibnamefont{Kurchan}}, \bibnamefont{and}
  \bibinfo{editor}{\bibfnamefont{J.}~\bibnamefont{Dalibard}}
  (\bibinfo{publisher}{Springer-Verlag}, \bibinfo{address}{Les Houches,
  France}, \bibinfo{year}{2003}).

\bibitem[{\citenamefont{Papadimitriou}(1991)}]{Papadimitriou}
\bibinfo{author}{\bibfnamefont{C.}~\bibnamefont{Papadimitriou}}, in
  \emph{\bibinfo{booktitle}{Proceedings of the 32th Annual Symposium on
  Foundations of Computer Science}} (\bibinfo{year}{1991}), pp.
  \bibinfo{pages}{163--169}.

\bibitem[{\citenamefont{Motwani and Ravaghan}(1995)}]{random_algo}
\bibinfo{author}{\bibfnamefont{R.}~\bibnamefont{Motwani}} \bibnamefont{and}
  \bibinfo{author}{\bibfnamefont{P.}~\bibnamefont{Ravaghan}},
  \emph{\bibinfo{title}{Randomized algorithms}} (\bibinfo{publisher}{Cambridge
  University Press}, \bibinfo{address}{Cambridge}, \bibinfo{year}{1995}).

\bibitem[{\citenamefont{Sch{\"o}ning}(2002)}]{schoning}
\bibinfo{author}{\bibfnamefont{U.}~\bibnamefont{Sch{\"o}ning}},
  \bibinfo{journal}{Algorithmica} \textbf{\bibinfo{volume}{32}},
  \bibinfo{pages}{615} (\bibinfo{year}{2002}), ISSN \bibinfo{issn}{0178-4617
  (print), 1432-0541 (electronic)}.

\bibitem[{\citenamefont{Baumer and Schuler}(2004)}]{schoning2}
\bibinfo{author}{\bibfnamefont{S.}~\bibnamefont{Baumer}} \bibnamefont{and}
  \bibinfo{author}{\bibfnamefont{R.}~\bibnamefont{Schuler}},
  \bibinfo{journal}{Lecture Notes in Computer Science}
  \textbf{\bibinfo{volume}{2919}}, \bibinfo{pages}{150} (\bibinfo{year}{2004}).

\bibitem[{\citenamefont{Semerjian and Monasson}(2003)}]{ws1}
\bibinfo{author}{\bibfnamefont{G.}~\bibnamefont{Semerjian}} \bibnamefont{and}
  \bibinfo{author}{\bibfnamefont{R.}~\bibnamefont{Monasson}},
  \bibinfo{journal}{Phys. Rev. E} \textbf{\bibinfo{volume}{67}},
  \bibinfo{pages}{066103} (\bibinfo{year}{2003}).

\bibitem[{\citenamefont{Barthel et~al.}(2003)\citenamefont{Barthel, Hartmann,
  and Weigt}}]{ws2}
\bibinfo{author}{\bibfnamefont{W.}~\bibnamefont{Barthel}},
  \bibinfo{author}{\bibfnamefont{A.~K.} \bibnamefont{Hartmann}},
  \bibnamefont{and} \bibinfo{author}{\bibfnamefont{M.}~\bibnamefont{Weigt}},
  \bibinfo{journal}{Phys. Rev. E} \textbf{\bibinfo{volume}{67}},
  \bibinfo{pages}{066104} (\bibinfo{year}{2003}).

\bibitem[{\citenamefont{Liggett}(1985)}]{cp}
\bibinfo{author}{\bibfnamefont{T.~M.} \bibnamefont{Liggett}},
  \emph{\bibinfo{title}{Interacting particle systems}}
  (\bibinfo{publisher}{Springer}, \bibinfo{address}{Berlin},
  \bibinfo{year}{1985}).

\bibitem[{\citenamefont{Alekhnovich and Ben-Sasson}(2006)}]{alek}
\bibinfo{author}{\bibfnamefont{M.}~\bibnamefont{Alekhnovich}} \bibnamefont{and}
  \bibinfo{author}{\bibfnamefont{E.}~\bibnamefont{Ben-Sasson}},
  \bibinfo{journal}{SIAM Journal on Computing} \textbf{\bibinfo{volume}{36}},
  \bibinfo{pages}{1248} (\bibinfo{year}{2006}).

\bibitem[{\citenamefont{Selman et~al.}(1994)\citenamefont{Selman, Kautz, and
  Cohen}}]{skc}
\bibinfo{author}{\bibfnamefont{B.}~\bibnamefont{Selman}},
  \bibinfo{author}{\bibfnamefont{H.~A.} \bibnamefont{Kautz}}, \bibnamefont{and}
  \bibinfo{author}{\bibfnamefont{B.}~\bibnamefont{Cohen}}, in
  \emph{\bibinfo{booktitle}{Proceedings of the Twelfth National Conference on
  Artificial Intelligence ({AAAI}'94)}} (\bibinfo{address}{Seattle},
  \bibinfo{year}{1994}), pp. \bibinfo{pages}{337--343}.

\bibitem[{\citenamefont{McAllester et~al.}(1997)\citenamefont{McAllester,
  Selman, and Kautz}}]{skc2}
\bibinfo{author}{\bibfnamefont{D.}~\bibnamefont{McAllester}},
  \bibinfo{author}{\bibfnamefont{B.}~\bibnamefont{Selman}}, \bibnamefont{and}
  \bibinfo{author}{\bibfnamefont{H.}~\bibnamefont{Kautz}}, in
  \emph{\bibinfo{booktitle}{Proceedings of the Fourteenth National Conference
  on Artificial Intelligence ({AAAI}'97)}} (\bibinfo{address}{Providence, Rhode
  Island}, \bibinfo{year}{1997}), pp. \bibinfo{pages}{321--326}.

\bibitem[{\citenamefont{Seitz et~al.}(2005)\citenamefont{Seitz, Alava, and
  Orponen}}]{rrt}
\bibinfo{author}{\bibfnamefont{S.}~\bibnamefont{Seitz}},
  \bibinfo{author}{\bibfnamefont{M.}~\bibnamefont{Alava}}, \bibnamefont{and}
  \bibinfo{author}{\bibfnamefont{P.}~\bibnamefont{Orponen}},
  \bibinfo{journal}{Journal of Statistical Mechanics: Theory and Experiment}
  \textbf{\bibinfo{volume}{2005}}, \bibinfo{pages}{P06006}
  (\bibinfo{year}{2005}).

\bibitem[{\citenamefont{Ardelius and Aurell}(2006)}]{asat}
\bibinfo{author}{\bibfnamefont{J.}~\bibnamefont{Ardelius}} \bibnamefont{and}
  \bibinfo{author}{\bibfnamefont{E.}~\bibnamefont{Aurell}},
  \bibinfo{journal}{Physical Review E (Statistical, Nonlinear, and Soft Matter
  Physics)} \textbf{\bibinfo{volume}{74}}, \bibinfo{eid}{037702}
  (pages~\bibinfo{numpages}{4}) (\bibinfo{year}{2006}).

\bibitem[{\citenamefont{Alava et~al.}(2007)\citenamefont{Alava, Ardelius,
  Aurell, Kaski, Krishnamurthy, Orponen, and Seitz}}]{circumspect}
\bibinfo{author}{\bibfnamefont{M.}~\bibnamefont{Alava}},
  \bibinfo{author}{\bibfnamefont{J.}~\bibnamefont{Ardelius}},
  \bibinfo{author}{\bibfnamefont{E.}~\bibnamefont{Aurell}},
  \bibinfo{author}{\bibfnamefont{P.}~\bibnamefont{Kaski}},
  \bibinfo{author}{\bibfnamefont{S.}~\bibnamefont{Krishnamurthy}},
  \bibinfo{author}{\bibfnamefont{P.}~\bibnamefont{Orponen}}, \bibnamefont{and}
  \bibinfo{author}{\bibfnamefont{S.}~\bibnamefont{Seitz}}
  (\bibinfo{year}{2007}), \bibinfo{note}{{\tt arXiv:0711.4902}}.

\bibitem[{\citenamefont{Chao and Franco}(1986)}]{Ch90_2}
\bibinfo{author}{\bibfnamefont{M.-T.} \bibnamefont{Chao}} \bibnamefont{and}
  \bibinfo{author}{\bibfnamefont{J.}~\bibnamefont{Franco}},
  \bibinfo{journal}{SIAM J. Comput.} \textbf{\bibinfo{volume}{15}},
  \bibinfo{pages}{1106} (\bibinfo{year}{1986}).

\bibitem[{\citenamefont{Chao and Franco}(1990)}]{Ch90_1}
\bibinfo{author}{\bibfnamefont{M.-T.} \bibnamefont{Chao}} \bibnamefont{and}
  \bibinfo{author}{\bibfnamefont{J.}~\bibnamefont{Franco}},
  \bibinfo{journal}{Inf. Sci.} \textbf{\bibinfo{volume}{51}},
  \bibinfo{pages}{289} (\bibinfo{year}{1990}).

\bibitem[{\citenamefont{Achlioptas}(2001)}]{Achltcs}
\bibinfo{author}{\bibfnamefont{D.}~\bibnamefont{Achlioptas}},
  \bibinfo{journal}{Theor. Comput. Sci.} \textbf{\bibinfo{volume}{265}},
  \bibinfo{pages}{159} (\bibinfo{year}{2001}).

\bibitem[{\citenamefont{Frieze and Suen}(1996)}]{Fr96}
\bibinfo{author}{\bibfnamefont{A.}~\bibnamefont{Frieze}} \bibnamefont{and}
  \bibinfo{author}{\bibfnamefont{S.}~\bibnamefont{Suen}}, \bibinfo{journal}{J.
  Algorithms} \textbf{\bibinfo{volume}{20}}, \bibinfo{pages}{312}
  (\bibinfo{year}{1996}).

\bibitem[{\citenamefont{Achlioptas et~al.}(2001)\citenamefont{Achlioptas,
  Kirousis, Kranakis, and Krizanc}}]{2pp_rigorous}
\bibinfo{author}{\bibfnamefont{D.}~\bibnamefont{Achlioptas}},
  \bibinfo{author}{\bibfnamefont{L.}~\bibnamefont{Kirousis}},
  \bibinfo{author}{\bibfnamefont{E.}~\bibnamefont{Kranakis}}, \bibnamefont{and}
  \bibinfo{author}{\bibfnamefont{D.}~\bibnamefont{Krizanc}},
  \bibinfo{journal}{Theor. Comput. Sci.} \textbf{\bibinfo{volume}{265}},
  \bibinfo{pages}{109} (\bibinfo{year}{2001}).

\bibitem[{\citenamefont{Cocco and Monasson}(2005)}]{Cocp}
\bibinfo{author}{\bibfnamefont{S.}~\bibnamefont{Cocco}} \bibnamefont{and}
  \bibinfo{author}{\bibfnamefont{R.}~\bibnamefont{Monasson}},
  \bibinfo{journal}{Ann. Math. Artif. Intell.} \textbf{\bibinfo{volume}{43}},
  \bibinfo{pages}{153} (\bibinfo{year}{2005}).

\bibitem[{\citenamefont{Deroulers and Monasson}(2004)}]{De04}
\bibinfo{author}{\bibfnamefont{C.}~\bibnamefont{Deroulers}} \bibnamefont{and}
  \bibinfo{author}{\bibfnamefont{R.}~\bibnamefont{Monasson}},
  \bibinfo{journal}{Europhysics Letters} \textbf{\bibinfo{volume}{68}},
  \bibinfo{pages}{153} (\bibinfo{year}{2004}).

\bibitem[{\citenamefont{Monasson}(2007)}]{Mo07}
\bibinfo{author}{\bibfnamefont{R.}~\bibnamefont{Monasson}}, in
  \emph{\bibinfo{booktitle}{Complex Systems}}, edited by
  \bibinfo{editor}{\bibfnamefont{J.~P.} \bibnamefont{Bouchaud}},
  \bibinfo{editor}{\bibfnamefont{M.}~\bibnamefont{M\'ezard}}, \bibnamefont{and}
  \bibinfo{editor}{\bibfnamefont{J.}~\bibnamefont{Dalibard}}
  (\bibinfo{publisher}{Elsevier}, \bibinfo{address}{Les Houches, France},
  \bibinfo{year}{2007}).

\bibitem[{\citenamefont{Cocco and Monasson}(2001)}]{Co01}
\bibinfo{author}{\bibfnamefont{S.}~\bibnamefont{Cocco}} \bibnamefont{and}
  \bibinfo{author}{\bibfnamefont{R.}~\bibnamefont{Monasson}},
  \bibinfo{journal}{Phys. Rev. Lett.} \textbf{\bibinfo{volume}{86}},
  \bibinfo{pages}{1654} (\bibinfo{year}{2001}).

\bibitem[{\citenamefont{Monasson}(2005)}]{Mon05}
\bibinfo{author}{\bibfnamefont{R.}~\bibnamefont{Monasson}},
  \emph{\bibinfo{title}{{A generating function method for the average-case
  analysis of DPLL.}}}, \bibinfo{howpublished}{{Lecture Notes in Computer
  Science 3624, 402-413 (2005).}} (\bibinfo{year}{2005}).

\bibitem[{\citenamefont{Beame et~al.}(2002)\citenamefont{Beame, Karp, Pitassi,
  and Saks}}]{beame}
\bibinfo{author}{\bibfnamefont{P.}~\bibnamefont{Beame}},
  \bibinfo{author}{\bibfnamefont{R.}~\bibnamefont{Karp}},
  \bibinfo{author}{\bibfnamefont{T.}~\bibnamefont{Pitassi}}, \bibnamefont{and}
  \bibinfo{author}{\bibfnamefont{M.}~\bibnamefont{Saks}},
  \bibinfo{journal}{SIAM Journal of Computing} \textbf{\bibinfo{volume}{31}},
  \bibinfo{pages}{1048} (\bibinfo{year}{2002}).

\bibitem[{\citenamefont{Tatikonda and Jordan}(2002)}]{TaJo}
\bibinfo{author}{\bibfnamefont{S.}~\bibnamefont{Tatikonda}} \bibnamefont{and}
  \bibinfo{author}{\bibfnamefont{M.}~\bibnamefont{Jordan}}, in
  \emph{\bibinfo{booktitle}{Proc. Uncertainty in Artificial Intell.}}
  (\bibinfo{year}{2002}), vol.~\bibinfo{volume}{18}, pp.
  \bibinfo{pages}{493--500}.

\bibitem[{\citenamefont{Montanari and Shah}(2007)}]{MoSh}
\bibinfo{author}{\bibfnamefont{A.}~\bibnamefont{Montanari}} \bibnamefont{and}
  \bibinfo{author}{\bibfnamefont{D.}~\bibnamefont{Shah}}, in
  \emph{\bibinfo{booktitle}{SODA}} (\bibinfo{year}{2007}), pp.
  \bibinfo{pages}{1255--1264}.

\bibitem[{\citenamefont{Braunstein and Zecchina}(2004)}]{BrZe}
\bibinfo{author}{\bibfnamefont{A.}~\bibnamefont{Braunstein}} \bibnamefont{and}
  \bibinfo{author}{\bibfnamefont{R.}~\bibnamefont{Zecchina}},
  \bibinfo{journal}{Journal of Statistical Mechanics: Theory and Experiment}
  \textbf{\bibinfo{volume}{2004}}, \bibinfo{pages}{P06007}
  (\bibinfo{year}{2004}).

\bibitem[{\citenamefont{Maneva et~al.}(2005)\citenamefont{Maneva, Mossel, and
  Wainwright}}]{Maneva}
\bibinfo{author}{\bibfnamefont{E.}~\bibnamefont{Maneva}},
  \bibinfo{author}{\bibfnamefont{E.}~\bibnamefont{Mossel}}, \bibnamefont{and}
  \bibinfo{author}{\bibfnamefont{M.~J.} \bibnamefont{Wainwright}}, in
  \emph{\bibinfo{booktitle}{SODA '05: Proceedings of the sixteenth annual
  ACM-SIAM symposium on Discrete algorithms}} (\bibinfo{publisher}{Society for
  Industrial and Applied Mathematics}, \bibinfo{address}{Philadelphia, PA,
  USA}, \bibinfo{year}{2005}), pp. \bibinfo{pages}{1089--1098}, ISBN
  \bibinfo{isbn}{0-89871-585-7}.

\bibitem[{\citenamefont{Aurell et~al.}(2004)\citenamefont{Aurell, Gordon, and
  Kirkpatrick}}]{AuKi}
\bibinfo{author}{\bibfnamefont{E.}~\bibnamefont{Aurell}},
  \bibinfo{author}{\bibfnamefont{U.}~\bibnamefont{Gordon}}, \bibnamefont{and}
  \bibinfo{author}{\bibfnamefont{S.}~\bibnamefont{Kirkpatrick}}, in
  \emph{\bibinfo{booktitle}{NIPS}} (\bibinfo{year}{2004}).

\bibitem[{\citenamefont{Parisi}(2003{\natexlab{a}})}]{Pa03}
\bibinfo{author}{\bibfnamefont{G.}~\bibnamefont{Parisi}}
  (\bibinfo{year}{2003}{\natexlab{a}}), \bibinfo{note}{{\tt
  arXiv:cs.CC/0301015}}.

\bibitem[{\citenamefont{Battaglia et~al.}(2004)\citenamefont{Battaglia,
  Kol\'a\ifmmode~\check{r}\else \v{r}\fi{}, and Zecchina}}]{SPy}
\bibinfo{author}{\bibfnamefont{D.}~\bibnamefont{Battaglia}},
  \bibinfo{author}{\bibfnamefont{M.}~\bibnamefont{Kol\'a\ifmmode~\check{r}\else
  \v{r}\fi{}}}, \bibnamefont{and}
  \bibinfo{author}{\bibfnamefont{R.}~\bibnamefont{Zecchina}},
  \bibinfo{journal}{Phys. Rev. E} \textbf{\bibinfo{volume}{70}},
  \bibinfo{pages}{036107} (\bibinfo{year}{2004}).

\bibitem[{\citenamefont{Chavas et~al.}(2005)\citenamefont{Chavas, Furtlehner,
  M\'{e}zard, and Zecchina}}]{reinforcement}
\bibinfo{author}{\bibfnamefont{J.}~\bibnamefont{Chavas}},
  \bibinfo{author}{\bibfnamefont{C.}~\bibnamefont{Furtlehner}},
  \bibinfo{author}{\bibfnamefont{M.}~\bibnamefont{M\'{e}zard}},
  \bibnamefont{and} \bibinfo{author}{\bibfnamefont{R.}~\bibnamefont{Zecchina}},
  \bibinfo{journal}{Journal of Statistical Mechanics: Theory and Experiment}
  \textbf{\bibinfo{volume}{2005}}, \bibinfo{pages}{P11016}
  (\bibinfo{year}{2005}).

\bibitem[{\citenamefont{Parisi}(2003{\natexlab{b}})}]{Pa03b}
\bibinfo{author}{\bibfnamefont{G.}~\bibnamefont{Parisi}}
  (\bibinfo{year}{2003}{\natexlab{b}}), \bibinfo{note}{{\tt
  arXiv:cond-mat/0308510}}.

\bibitem[{SP_()}]{SP_implementation}
\urlprefix\url{http://www.ictp.trieste.it/~zecchina/SP}.

\bibitem[{\citenamefont{Montanari et~al.}(2007)\citenamefont{Montanari,
  Ricci-Tersenghi, and Semerjian}}]{Allerton}
\bibinfo{author}{\bibfnamefont{A.}~\bibnamefont{Montanari}},
  \bibinfo{author}{\bibfnamefont{F.}~\bibnamefont{Ricci-Tersenghi}},
  \bibnamefont{and} \bibinfo{author}{\bibfnamefont{G.}~\bibnamefont{Semerjian}}
  (\bibinfo{year}{2007}), \bibinfo{note}{{\tt arXiv:0709.1667}, to be published
  in the Proceedings of the 45th Allerton Conference (2007)}.

\bibitem[{\citenamefont{Feige}(2002)}]{Feige1}
\bibinfo{author}{\bibfnamefont{U.}~\bibnamefont{Feige}}, in
  \emph{\bibinfo{booktitle}{STOC}} (\bibinfo{year}{2002}), pp.
  \bibinfo{pages}{534--543}.

\bibitem[{\citenamefont{Feige et~al.}(2006)\citenamefont{Feige, Mossel, and
  Vilenchik}}]{Feige2}
\bibinfo{author}{\bibfnamefont{U.}~\bibnamefont{Feige}},
  \bibinfo{author}{\bibfnamefont{E.}~\bibnamefont{Mossel}}, \bibnamefont{and}
  \bibinfo{author}{\bibfnamefont{D.}~\bibnamefont{Vilenchik}},
  \emph{\bibinfo{title}{{Complete convergence of message passing algorithms for
  some satisfiability problems.}}}, \bibinfo{howpublished}{{Lecture Notes in
  Computer Science 4110, 339-350 (2006).}} (\bibinfo{year}{2006}).

\bibitem[{\citenamefont{Altarelli et~al.}(2007)\citenamefont{Altarelli,
  Monasson, and Zamponi}}]{AMZ06}
\bibinfo{author}{\bibfnamefont{F.}~\bibnamefont{Altarelli}},
  \bibinfo{author}{\bibfnamefont{R.}~\bibnamefont{Monasson}}, \bibnamefont{and}
  \bibinfo{author}{\bibfnamefont{F.}~\bibnamefont{Zamponi}},
  \bibinfo{journal}{Journal of Physics A: Mathematical and Theoretical}
  \textbf{\bibinfo{volume}{40}}, \bibinfo{pages}{867} (\bibinfo{year}{2007}).

\bibitem[{\citenamefont{Coja-Oghlan et~al.}()\citenamefont{Coja-Oghlan,
  Krivelevich, and Vilenchik}}]{why}
\bibinfo{author}{\bibfnamefont{A.}~\bibnamefont{Coja-Oghlan}},
  \bibinfo{author}{\bibfnamefont{M.}~\bibnamefont{Krivelevich}},
  \bibnamefont{and}
  \bibinfo{author}{\bibfnamefont{D.}~\bibnamefont{Vilenchik}},
  \emph{\bibinfo{title}{Why almost all k-cnf formulas are easy}},
  \bibinfo{note}{to appear (2007)}.

\bibitem[{\citenamefont{Krzakala et~al.}(2004)\citenamefont{Krzakala, Pagnani,
  and Weigt}}]{col1}
\bibinfo{author}{\bibfnamefont{F.}~\bibnamefont{Krzakala}},
  \bibinfo{author}{\bibfnamefont{A.}~\bibnamefont{Pagnani}}, \bibnamefont{and}
  \bibinfo{author}{\bibfnamefont{M.}~\bibnamefont{Weigt}},
  \bibinfo{journal}{Phys. Rev. E} \textbf{\bibinfo{volume}{70}},
  \bibinfo{pages}{046705} (\bibinfo{year}{2004}).

\bibitem[{\citenamefont{Zdeborov\'{a} and Krzakala}(2007)}]{col2}
\bibinfo{author}{\bibfnamefont{L.}~\bibnamefont{Zdeborov\'{a}}}
  \bibnamefont{and} \bibinfo{author}{\bibfnamefont{F.}~\bibnamefont{Krzakala}},
  \bibinfo{journal}{Physical Review E (Statistical, Nonlinear, and Soft Matter
  Physics)} \textbf{\bibinfo{volume}{76}}, \bibinfo{eid}{031131}
  (pages~\bibinfo{numpages}{29}) (\bibinfo{year}{2007}).

\bibitem[{\citenamefont{Barthel et~al.}(2002)\citenamefont{Barthel, Hartmann,
  Leone, Ricci-Tersenghi, Weigt, and Zecchina}}]{Ba02}
\bibinfo{author}{\bibfnamefont{W.}~\bibnamefont{Barthel}},
  \bibinfo{author}{\bibfnamefont{A.~K.} \bibnamefont{Hartmann}},
  \bibinfo{author}{\bibfnamefont{M.}~\bibnamefont{Leone}},
  \bibinfo{author}{\bibfnamefont{F.}~\bibnamefont{Ricci-Tersenghi}},
  \bibinfo{author}{\bibfnamefont{M.}~\bibnamefont{Weigt}}, \bibnamefont{and}
  \bibinfo{author}{\bibfnamefont{R.}~\bibnamefont{Zecchina}},
  \bibinfo{journal}{Phys. Rev. Lett.} \textbf{\bibinfo{volume}{88}},
  \bibinfo{pages}{188701} (\bibinfo{year}{2002}).

\bibitem[{\citenamefont{Montanari and Semerjian}(2006{\natexlab{b}})}]{MoSe2}
\bibinfo{author}{\bibfnamefont{A.}~\bibnamefont{Montanari}} \bibnamefont{and}
  \bibinfo{author}{\bibfnamefont{G.}~\bibnamefont{Semerjian}},
  \bibinfo{journal}{J. Stat. Phys.} \textbf{\bibinfo{volume}{125}},
  \bibinfo{pages}{23} (\bibinfo{year}{2006}{\natexlab{b}}).

\end{thebibliography}

\end{document}